\documentclass{jfm}
\usepackage{url}
\usepackage{amsmath}
\usepackage{amssymb,amsbsy}
\usepackage[dvips]{graphicx}
\usepackage[mathscr]{euscript}

\ifCUPmtlplainloaded \else \checkfont{eurm10} \iffontfound
\IfFileExists{upmath.sty} {\typeout{^^JFound AMS Euler Roman fonts on
the system, using the 'upmath' package.^^J}%
       \usepackage{upmath}} {\typeout{^^JFound AMS Euler Roman fonts
       on the system, but you dont seem to have the}%
        \typeout{'upmath' package installed.  JFM.cls can take
        advantage of these fonts,^^Jif you use 'upmath' package.^^J}%
       } \else %
     \fi
\fi

\ifCUPmtlplainloaded \else \checkfont{msam10} \iffontfound
\IfFileExists{amssymb.sty} {\typeout{^^JFound AMS Symbol fonts on the
system, using the 'amssymb' package.^^J}%
          \usepackage{amssymb}%
          
         }{} \fi \fi

\ifCUPmtlplainloaded \else \IfFileExists{amsbsy.sty}
{\typeout{^^JFound the 'amsbsy' package on the system, using it.^^J}%
      \usepackage{amsbsy}}
      {\providecommand\boldsymbol[1]{\mbox{\boldmath $##1$}}} \fi

\newcommand{\W}{\Omega} 

\newsavebox{\astrutbox} \sbox{\astrutbox}{\rule[-5pt]{0pt}{20pt}}

\newcommand\etal{\mbox{\textit{et al.\ }}}

\title[Anisotropy and cyclone-anticyclone asymmetry in decaying rotating turbulence]{Anisotropy and cyclone-anticyclone asymmetry in decaying rotating turbulence}

\author[Moisy \etal]{F.\ns M\ls O\ls I\ls S\ls Y$^{1}$, \ns
C.\ns M\ls O\ls R\ls I\ls Z\ls E$^{1}$, \ns M.\ns R\ls A\ls B\ls A\ls
U\ls D$^{1}$, \ns J.\ns S\ls O\ls M\ls M\ls E\ls R\ls I\ls A$^2$}

\affiliation{$^{1}$Universit\'e Paris-Sud 11, Universit\'e Pierre et Marie Curie, CNRS; Laboratoire FAST, B\^at. 502, F-91405 Orsay, France\\
$^{2}$Coriolis/LEGI, 21 avenue des Martyrs, F-38 000 Grenoble, France}

\pubyear{2009} \volume{1} \pagerange{1--31} \date{\today}
\begin{document}

\maketitle

\begin{abstract}

The effect of a background rotation on the decay of homogeneous turbulence produced by a grid is experimentally investigated. Experiments have been performed in a channel mounted in the large-scale 'Coriolis' rotating platform, and measurements have been carried out in the planes normal and parallel to the rotation axis using particle image velocimetry. After a short period of about 0.4 tank rotation where the energy decays as $t^{-6/5}$, as in classical isotropic turbulence, the energy follows a shallower decay law compatible with  $t^{-3/5}$, as dimensionally expected for energy transfers governed by the linear timescale $\Omega^{-1}$. The crossover occurs at a Rossby number $Ro \simeq 0.25$, without noticeable dependence with the grid Rossby number. After this transition, anisotropy develops in the form of vertical layers where the initial vertical velocity remains trapped.  These layers of nearly constant vertical velocity become thinner as they are advected and stretched by the large-scale horizontal flow, producing significant horizontal gradient of vertical velocity which eventually become unstable. After the $Ro \simeq 0.25$ transition, the vertical vorticity field first develops a cyclone-anticyclone asymmetry, reproducing the growth law of the vorticity skewness,  $S_\omega(t) \simeq (\Omega t)^{0.7}$, reported by Morize, Moisy \& Rabaud [{\it Phys. Fluids} {\bf 17} (9), 095105 (2005)]. At larger time, however, the vorticity skewness decreases and eventually returns to zero. The present results indicate that the shear instability of the vertical layers contribute significantly to the re-symmetrisation of the vertical vorticity at large time, by re-injecting vorticity fluctuations of random sign at small scales. These results emphasize the importance of the initial conditions in the decay of rotating turbulence.

\end{abstract}

\section{Introduction}
\label{sec:intro}

Turbulence subjected to solid body rotation is a problem of first importance for engineering, geophysical and astrophysical flows. Its dynamics is dictated by a competition between linear and non-linear effects. Linear effects, driven by the Coriolis force, include anisotropic propagation of energy by inertial waves, preferentially along the rotation axis (hereafter called 'vertical' axis by convention), on the timescale of the system rotation $\Omega^{-1}$ (Greenspan 1968).  Nonlinear interactions, on the other hand, are responsible for energy transfers towards 'horizontal' modes (Jacquin \etal 1990; Waleffe 1993). For infinite rotation rate, i.e. for vanishing Rossby number, these inertial waves reduce to Taylor-Proudman columns, corresponding to a two-dimensional flow invariant along the rotation axis. Importantly, this two-dimensional flow is not two-component in general, because the third (vertical) velocity component, insensitive to the Coriolis force, behaves as a passive scalar field transported by the horizontal flow. This 'passive' vertical velocity, originating from the initial conditions in the case of decaying turbulence, may however become 'active' through shear instabilities at small scale. This mechanism may have considerable importance in the nature of the decay and the partial two-dimensionalisation of an initially 3D turbulence subjected to background rotation.


The present paper reports an experimental study of the influence of the background rotation on the decay of an initially isotropic turbulence. Turbulence is generated by translating a grid in a channel mounted on the large-scale 'Coriolis'  rotating platform.  The aim of this paper is first to characterize in details the decay law of the energy, in a situation where the effects of the lateral confinement can be neglected.   This situation contrasts with the previous experiments by Morize, Moisy \& Rabaud (2005) and Morize \& Moisy (2006a), performed in a rotating tank with an aspect ratio of order 1, showing significant confinement effects. Second, the anisotropy growth is investigated, with the aim to characterize the influence at large time of the initial vertical fluctuations on the vertical vorticity statistics.

For turbulence subjected to moderate rotation (Rossby number $Ro \simeq O(1)$), the linear and nonlinear time scales are of the same order, resulting in a complex interplay between linear energy propagation by inertial waves and anisotropic energy transfers by nonlinear interactions.  This complexity is unavoidable in decaying rotating turbulence starting  at large Rossby number, in which the instantaneous Rossby number decays and crosses $O(1)$ at some transition time. At this time the effects of the rotation,  namely the  anisotropy growth and the cyclone-anticyclone symmetry breaking, become significant and accumulate as time proceeds. Accordingly, the statistical properties of rotating turbulence at large time are the result of the turbulence history integrated from the initial state, and may therefore depend on the details of the initial state. Generic properties should however be expected if the initial state is 3D isotropic turbulence with $Ro \gg 1$, which is the situation examined in the present paper.

Because of the fast growth of the vertical correlation due to inertial wave propagation (Jacquin \etal 1990; Squires \etal 1994), confinement along the vertical axis plays a significant role in the dynamics of rotating turbulence. This vertical confinement is usually present either through solid or free-slip boundaries in experiments, or through periodic boundary conditions in numerical simulations. In all cases, comparisons with homogeneous turbulence in idealized unbounded systems should be made carefully. One consequence of the vertical confinement by solid or free-slip boundaries is to preferentially align the axis of the vortices normal to the walls, therefore reinforcing the two-dimensional nature of the large scales, as observed by Hopfinger, Browand \& Gagne (1982) and Godeferd \& Lollini (1999). Second, confinement selects a set of discrete resonant inertial modes (Bewley \etal 2007), which may couple to the small-scale turbulence. Third, an extra mechanism of dissipation of the inertial waves take place in the boundary layers, acting on the Ekman time scale $h (\nu \Omega)^{-1/2}$, where $h$ is the confinement scale along the rotation axis, which may dominate the energy decay at large time (Phillips 1963; Ibbetson \& Tritton 1982; Morize \& Moisy 2006a).

The most remarkable feature of rotating turbulence is the spontaneous emergence of long-lived columnar vortices aligned with the rotation axis (Hopfinger \etal 1982; Smith \& Waleffe 1999; Longhetto \etal 2002).  Both linear and non-linear mechanisms have been proposed to explain the formation of those columnar structures (Cambon \& Scott 1999; Cambon 2001; Davidson, Staplehurst \& Dalziel 2006; Staplehurst, Davidson \& Dalziel 2008).  Moreover, a symmetry breaking between cyclonic and anticyclonic vortices is observed, both in forced (Hopfinger \etal 1982) and decaying (Bartello, M\'etais \& Lesieur 1994; Smith \& Waleffe 1999) turbulence. This symmetry breaking has received considerable interest in recent years. It  has been quantified in terms of the vorticity skewness, $S_\omega = {\langle \omega_z^3 \rangle}/{\langle \omega_z^2 \rangle^{3/2}}$ (where $\omega_z$ is the vorticity component along the rotation axis), which is found to be positive for $Ro \simeq 1$ (Bartello \etal 1994; Morize \etal 2005; Rupper-Felsot \etal 2005; Bourouiba \& Bartello 2007; Bokhoven \etal 2008; Staplehurst \etal 2008). In decaying rotating turbulence, starting from initial conditions such that $Ro \gg 1$, a power-law growth has been observed in the form $S_\omega \simeq (\Omega t)^{0.6 \pm 0.1}$ by Morize \etal (2005). The fact that the time appears through the non-dimensional combination $\Omega t$ actually suggests a build-up of vorticity skewness acting on the linear time scale $\Omega^{-1}$.

Several explanations have been proposed for the cyclone-anticyclone asymmetry growth, although none provides a complete explanation of the experimental data. First, in a rotating frame, the vortex stretching term for the axial vorticity, $(2\Omega + \omega_z) \partial u_z / \partial z$, is larger for cyclonic than for anticyclonic vorticity.  \cite{Gence01} have shown that, for isotropic turbulence suddenly subjected to a background rotation, $S_\omega$ grows linearly at short time, i.e. for $t \ll \Omega^{-1}$. At larger time the growth is expected to be slower, because the vertical strain $\partial u_z / \partial z$ is reduced by the rotation. Second, anticyclonic vortices are more prone to centrifugal instabilities. This effect can be readily shown for idealized axisymmetric vortices, for which the generalized Rayleigh criterion in a rotating frame (Kloosterziel \& van Heijst 1991), $\phi(r) = 2 (\omega_z + 2 \Omega)(u_\theta/r + \Omega)$, is more likely to become negative for anticyclonic vorticity. Sreenivasan \& Davidson (2008) have shown, using a model of axisymmetric vortex patches, that cyclonic vortices first develop columnar structures, while anticyclonic vortices become centrifugally unstable.

More surprisingly, a decrease of $S_\omega$ has been reported at large times (smaller $Ro$) by Morize \etal (2005, 2006b), and later confirmed by Bokhoven \etal (2008). There is no general agreement concerning this unexpected return to vorticity symmetry. Several two-dimensional mechanisms, such as cyclone merging or diffusion, may induce a reduction of this quantity. The role of the confinement was also suggested  by Morize \etal (2005), nameling the non-linear Ekman pumping on the rigid walls. However it was questioned by Bokhoven \etal (2008), who also reported a decrease of $S_\omega$ at large times, but in a numerical simulation with periodic boundary conditions, hence with no Ekman pumping.  The non-monotonic time evolution of the vorticity skewness is confirmed by the present experiment, although the present boundary conditions significantly differs from those of Morize \etal (2005). The present results suggest another contribution for this decrease at large times: As time proceeds, the vertical velocity, initiated by the 3D initial conditions, forms vertically coherent layers, either ascending or descending, passively transported by the large-scale quasi-two-dimensional flow. The straining of those layers by the horizontal motion produces smaller scales, in a process similar to the enstrophy cascade in two-dimensional turbulence. This mechanism reinforces the vertical shear, making those layers prone to inertial instabilities, producing small-scale horizontal vorticity, which in turn produces either cyclonic or anticyclonic vertical vorticity, resulting in a reduction of $S_\omega$ at large times. This re-injection of symmetric vorticity fluctuations at small scale is thought to be a generic mechanism in decaying rotating turbulence, provided the initial state contains a significant amount of vertical velocity, which is the case for an initial 3D isotropic turbulence.

The paper is organised as follows. Section~\ref{sec:setup} describes the experimental setup and the PIV measurements. In Sec.~\ref{sec:lsp} the large scale flows and the procedure to extract the turbulent flow component are detailed. The influence of the background rotation on the energy decay and the time evolution of the non-dimensional numbers are presented in Sec.~\ref{sec:decay}.  The anisotropy growth and the formation of the vertical layers are characterised in Sec.~\ref{sec:aniso}. The structure and dynamics of the vertical vorticity field is described in Sec.~\ref{sec:vor}, with emphasis given on the cyclone-anticyclone asymmetry growth and its decay induced by the shear instability of the vertical layers. Finally, Sec.~\ref{sec:concl} summarizes the different regimes observed during the decay.

\section{Experimental set-up and procedure}
\label{sec:setup}

\subsection{Experimental apparatus}

\begin{figure}
\begin{center}
\includegraphics[width=11cm]{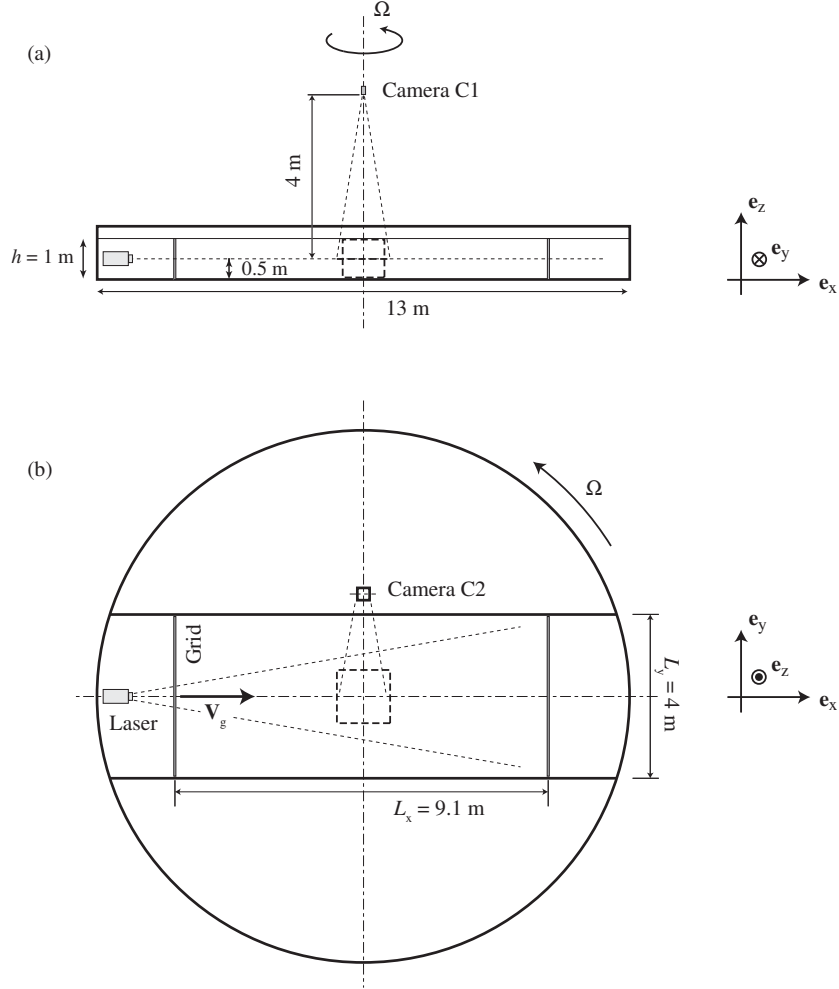}
\caption{Side view (a) and top view (b) of the experimental setup. The grid is translated from left to right, along ${\bf e}_x$. The angular velocity is ${\bf \Omega} = \Omega {\bf e}_z$, with $\Omega > 0$ (anticlockwise rotation). The PIV camera is located either at C1 or C2,
for measurements in the horizontal and vertical planes, respectively. The dashed squares show the corresponding imaged areas.} \label{fig:setup}
\end{center}
\end{figure}

The experimental setup, sketched in figure~\ref{fig:setup},
consists in a 13~m~$\times$~4~m water channel mounted on the
`Coriolis' rotating platform. Details about the rotating
platform may be found in Praud \etal (2005, 2006) and Praud \&
Fincham (2005), and only the features specific to the present
experiments are described here. The whole tank is filled with
water to avoid leaks from the channel, with a depth of
$h=1$~m. One experiment without rotation, and 3 experiments with
rotation periods of $T = 30$, 60 and 120~s, have been carried out (see
table~\ref{tab:fp}). The angular velocity $\Omega = 2\pi / T$ is constant within a precision of $\Delta \Omega / \Omega < 10^{-4}$. The parabolic deformation of the surface height induced by the rotation is 0.3~cm (resp.
4.5~cm) for the lowest (resp. highest) rotation rate.

The turbulence is generated by horizontally translating a vertical grid at a constant velocity $V_g = 30$~cm~s$^{-1}$ along the channel over a distance of $L_x = 9.1$~m (see Movie 1).  The grid is made of square bars of 30~mm, with a mesh size of $M =140$~mm and a solidity ratio of 0.38. The grid cross section is 4~m~$\times$~1~m~$=28M \times 7M$, ensuring negligible lateral confinement effects. It is hung from a carriage moving above the free surface, and its displacement is driven through a computer-controlled DC motor,
ensuring good repeatability of the experiments. The velocity of
the grid increases linearly from 0 to $V_g$ and, remains constant
in the central part, and decreases linearly back to zero at the end of the channel. The acceleration and deceleration are chosen to minimize the generation of gravity waves by accumulation of fluid between the grid and the
endwalls of the channel. The time at which the grid reach the center of the channel, where the measurements are performed, defines the origin $t=0$.  The streamwise, spanwise and vertical axis are noted ${\bf e}_x$, ${\bf e}_y$ and ${\bf e}_z$ respectively.  

The initial condition of an experiment is characterized by two
nondimensional parameters, the Reynolds and Rossby numbers based
on the grid velocity and grid mesh,
$$
Re_g = \frac{V_g M}{\nu}, \qquad Ro_g = \frac{V_g}{2 \Omega M},
$$
where $\nu$ is the water kinematic viscosity. The grid Reynolds number is
constant for all the experiments, $Re_g = 4.20 \times 10^4$, while the Rossby number lies in the range $5.1-20.4$ (table~\ref{tab:fp}).  Accordingly, the turbulent energy production in the near wake of the grid is expected to be weakly affected by the rotation, so the initial state can be considered as approximately isotropic.

\begin{table}
\begin{center}
\begin{tabular}{p{6cm}p{1.1cm}p{1.1cm}p{1.1cm}p{1.1cm}}
\hline \hline
Rotation period $T$ (s) & - & 120& 60& 30 \\
Symbol & $\star$ & $\circ$ & $\square$ & $\triangle$ \\
Angular velocity $\Omega$ (rad s$^{-1}$) & 0 & 0.052 & 0.105 & 0.209 \\
Grid Rossby number $Ro_g$ & - & 20.4 & 10.2 & 5.1 \\
Ekman layer thickness $\delta_E$ (mm) & - & 4.4 & 3.1 & 2.2 \\
Ekman time scale $t_E$ (s) & - & 4370 & 3090 & 2185 \\
\hline \hline
\end{tabular}
\caption{Flow parameters. The symbols are used in the following figures.}
\label{tab:fp}
\end{center}
\end{table}

\subsection{Particle Image Velocimetry measurements}
\label{sec:meas}

\subsubsection{PIV setup}

A high-resolution particle image velocimetry system, based on a 14
bits $2048 \times 2048$ pixels camera (PCO.2000), was used in these
experiments. Water was seeded by Chemigum P83 particles,
250~$\mu$m in diameter, and illuminated by a 8W laser sheet of
thickness 1~cm.  The laser sheet was transmitted to the measurement area through an immersed mirror at 45$^\mathrm{o}$ at one end of the channel.

Two types of measurements were performed:

\begin{enumerate}

\item in a centered square area of 1.3~m $\times$ 1.3~m in the
horizontal plane $({\bf e}_x, {\bf e}_y)$ at mid-height
($z=0.5$~m). The camera is located 4~m above the horizontal
laser sheet (C1 in figure~\ref{fig:setup}), and the area is imaged through the
free surface (see Sec.~\ref{sec:fsd}).

\item in a 1.1~m $\times$ 1~m area in the vertical plane $({\bf
e}_x, {\bf e}_z)$ in the middle of the channel. The
plane is imaged through a window in the lateral wall  (C2 in figure~\ref{fig:setup}), so that the measurements
are not affected by free surface disturbances.

\end{enumerate}

Up to six decay experiments of 1 hour (7700 grid timescales $M/V_g$) have been carried out for each rotation rate and, for each decay, 400 image pairs are recorded. Since the characteristic velocity decreases in time, the delay between the two successive images of a pair is made to gradually increase during the acquisition sequence, from 125~ms to 2~s, so that the typical particles displacement remains approximately constant throughout the decay. The time delay between image pairs is also gradually increased during the decay, from 2~s to 20~s. The results are ensemble-averaged over the $N_r=6$ realizations in the horizontal plane, and $N_r=4$ in the vertical plane. Although this is enough to achieve statistical convergence at small times, when the correlation length is significantly smaller than the imaged area, the convergence becomes questionable at large times, when the imaged area contains in average 1 large-scale structure or less.

\subsubsection{PIV resolution}
\label{sec:pivres}

The PIV computations have been performed using the software Davis\footnote{LaVision GmbH, Anna-Vandenhoeck-Ring 19, D-37081 Goettingen, Germany.}, and
the statistical analysis of the velocity fields using the  PivMat toolbox under Matlab\footnote{\url{http://www.fast.u-psud.fr/pivmat}}. Interrogation windows of size $32 \times
32$ pixels, with an overlap of 16~pixels, were used.  The final
velocity fields are defined on a $128 \times 128$ grid. The
spatial resolution is $\Delta x = 10$~mm, which is of the order of the laser
sheet thickness.  With a particle displacement resolution of 0.1~pixel,
achieved using a classical subpixel interpolation scheme for the
correlation function, a velocity signal-to-noise ratio of $2 \times 10^{-2}$ is obtained.

Due to the moderate spatial resolution of 10~mm, the velocity field inside the Ekman boundary layer, of thickness $\delta_E = (\nu / \Omega)^{1/2} \simeq 2.2 - 4.4$~mm (see Table~\ref{tab:fp}), cannot be resolved. Assuming isotropy in the bulk of the flow, which is valid only in the non-rotating case or at small time, the smallest turbulent scale can be estimated by the Kolmogorov scale $\eta = (\nu^3 / \epsilon)^{1/4}$, where the dissipation rate $\epsilon$ can be computed from the energy decay, $\epsilon \simeq  - (3/2) \partial (u'_x)^2 / \partial t$ (energy decays are detailed in Sec.~\ref{sec:decay}). The scale $\eta$ is of order of 0.4~mm $ \simeq \Delta x / 30$ at $t \simeq 20 M/V_g$, for all rotation rates, so that the smallest scales are not resolved at the beginning of the decay. Accordingly, the measured vorticity at scale $\Delta x$ underestimates the actual one. Vorticity measurements become reliable when $\eta > \Delta x / 5$, which is satisfied for $t > 250 M/V_g$ only. At the end of the decay, $\eta$ is of order of 8~mm for $\Omega = 0$, in which case the vorticity can be accurately computed from the PIV measurements.

\subsection{Free surface disturbances}
\label{sec:fsd}

When the flow is imaged from above, a significant uncertainty
originates from the refraction through disturbances at the free
surface. These perturbations induce an additional apparent particle
displacement, $\delta {\bf x}_\mathrm{FSD} = (1-1/n_w) \, (h/2) \, {\bf
\nabla} h$, with $n_w$ the water refraction index, ${\bf \nabla} h$ the surface gradient, and $h/2$ the path length of the refracted light rays
(see e.g. Moisy, Rabaud \& Salsac 2009).

Two sources of free surface disturbances of small wavelengths have been identified: perturbations due to the wake of the grid, which prevent the measurements during typically 10~s ($\simeq 20 M/V_g$) after the grid translation, and surface waves continuously excited by residual mechanical vibrations from the
platform structure, in the range 5-10~Hz (wavelength of 10-30~cm),
which may be non negligible for the highest rotation rate. The
resulting velocity contamination is of order of $|{\bf u}_\mathrm{FSD}| = |\delta
{\bf x}_\mathrm{FSD}| / \delta t$, with $\delta t$ the inter-frame
time.  The disturbance amplitude was significantly reduced when a thin layer of impurities (dust) was present at the water surface. In practice, the velocity contamination originating from those
free surface disturbances can be neglected when computing quantities based on the velocity field itself, or on the antisymmetric part of the velocity
derivative tensor (e.g. vorticity), but may significantly affect
its symmetric part (e.g. horizontal divergence).

\section{Large scale properties of the flow}
\label{sec:lsp}

\subsection{Flow visualisations}

\begin{figure}
\begin{center}
\includegraphics[width=9.5cm]{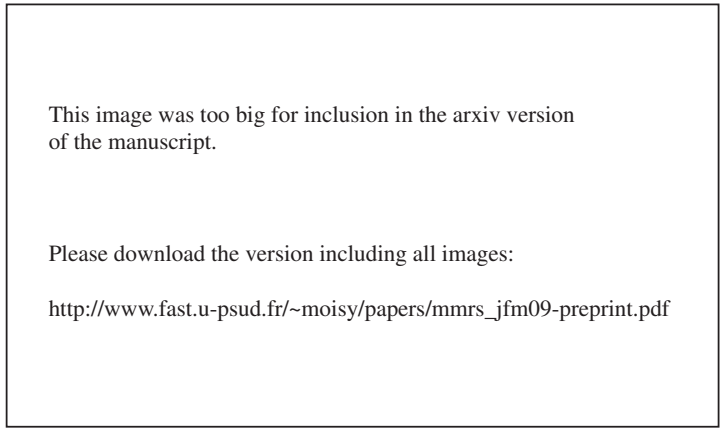}
\caption{Horizontal and vertical snapshots of the velocity fields taken at $t = 360$~s~$\simeq 770 M/V_g$ after the grid translation, without rotation (left figures) and with rotation at $\Omega = 0.10$~rad~s$^{-1}$. The imaged area is $1 \times 1$~m. The color shows the corresponding normal component of the vorticity, $\omega_z(x,y)$ and $\omega_y(x,z)$. For the rotating cases, this time corresponds to 6 tank rotations. The color palette for the vorticity ranges from $-0.1$ to $0.1$~rad~s$^{-1}$. Note that mean flow in the direction of the grid velocity (along ${\bf e}_x$, which is present for $\Omega = 0$ but inhibited by the background rotation for $\Omega \neq 0$.}
\label{fig:snap}
\end{center}
\end{figure}

First insight into the influence of the background rotation on the turbulence decay may be obtained by comparing the horizontal and vertical vorticity fields shown in figure~\ref{fig:snap}, in the non-rotating case (left) and in an experiment rotating at $\Omega = 0.10$~rad~s$^{-1}$ (right). Those snapshots are obtained 360~s after the grid translation ($770 M/V_g$). At this time, the turbulent Reynolds number (defined in Sec.~\ref{sec:rero}) is 400 and 700 for the non-rotating and rotating cases respectively, and the Rossby number for the rotating case is 0.06.

While the vorticity fields $\omega_z$ and $\omega_y$ for the non-rotating cases are similar in the two measurement planes, as expected for approximately isotropic turbulence, they strongly differ in the rotating case. Movies 2 and 3 (see supplementary matarial) of $\omega_z$ and $\omega_y$  clearly show the two essential features of the turbulenc decay in the rotating frame, namely the anisotropy growth and the cyclone-anticyclone asymmetry.

The vertical vorticity $\omega_z$ shows strong large-scale vortices, mostly cyclonic (in red), surrounded by shear layers. In the vertical plane, the spanwise vorticity $\omega_y$ shows elongated structures of alternating sign, originating from layers of ascending and descending fluid. The dominant contribution of $\omega_y$ comes from the vertical shear, $\partial u_z / \partial x$, except near the top and bottom boundary layers where the horizontal shear $\partial u_x / \partial z$ is dominant. Sections~\ref{sec:aniso} and \ref{sec:vor} will be devoted to a more detailed characterisation of the structure of the flow.

\subsection{Large scale flows}
\label{sec:meanflow}

Translating a grid in a closed volume is ideally designed to produce homogeneous turbulence with zero mean flow. However, repeatable flow features are found over successive realisations, so a careful separation between mean and turbulent flows is necessary to analyse the present experiments.

\begin{figure}
\begin{center}
\includegraphics[width=8cm]{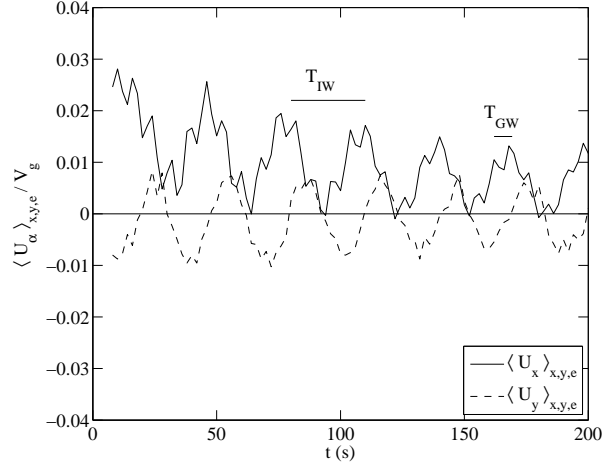}
\caption{Time evolution of the ensemble and spatially
averaged streamwise (---) and spanwise (- -)
velocity components, for  $\Omega = 0.1$~rad~s$^{-1}$ (rotation period $T = 60$~s).
The fast oscillation, of period $T_{GW} \simeq 7.3$~s, is a longitudinal gravity wave, and the slow oscillation, of period $T_{IW} = T/2 = 30$~s, is an anticyclonic inertial wave.} 
\label{fig:mf}  
\end{center}                                            
\end{figure}

The main repeatable features of the flow generated by the grid translation can be inferred from figure~\ref{fig:mf}, where the spatially- and ensemble-averaged streamwise, $\langle U_x \rangle$, and spanwise, $\langle U_y \rangle$, velocity components are shown as a function of time, for $\Omega = 0.1$~rad~s$^{-1}$ (the averaging procedure is detailed in Sec.~\ref{sec:ave}).  A large scale circulation (LSC) is clearly visible, traced here as a systematic positive, slowly decaying, streamwise velocity. In addition, two oscillatory flows, namely a gravity wave (GW) and an inertial wave (IW), are found to superimpose to the LSC.

{\it Gravity wave} (GW): As the grid is translated along the channel, it pushes a significant amount of water near the endwall, which initiates a fast longitudinal gravity wave. This sloshing mode is present both in the rotating and non-rotating cases. Its period $T_\mathrm{GW}$ is given by $\lambda / c$, where the wavelength $\lambda$ is twice the channel length and the phase velocity in the shallow-water approximation is $c=(gh)^{1/2} \simeq 3.13$~m~s$^{-1}$, yielding $T_\mathrm{GW} \simeq 7.3$~s (see figure~\ref{fig:mf}).

{\it Large scale circulation} (LSC): This mean flow is responsible for the positive mean streamwise velocity at mid-height in figure~\ref{fig:mf}, of initial amplitude of order $\simeq 2 \times 10^{-2} V_g$. It originates from the boundary condition asymmetry between the solid boundary at $z=0$ and the free surface at $z=h$, yielding a significant residual horizontal shear $\partial \langle U_x \rangle / \partial z > 0$, of initial amplitude $\simeq 0.03 V_g / h \simeq 10^{-2}$~s$^{-1}$.  This shear flow is indeed visible in figure~\ref{fig:snap}(c) for $\Omega=0$.  This positive streamwise flow is also present in the depth-averaged velocity measured in the vertical plane, and must be compensated by a horizontal recirculation flow along the lateral walls.

{\it Inertial wave} (IW): When rotation is present, the mean horizontal shear produced by the grid further excites an inertial wave, of period half the rotation period of the tank $T_\mathrm{IW} = T/2$ ($T_\mathrm{IW} = 30$~s in figure~\ref{fig:mf}). Visualisations in the vertical plane show that the
IW can be approximately described as the oscillating shearing motion of horizontal layers, of thickness $h/2$, so that the associated wavenumber is vertical. In the horizontal plane, the signature of this IW is a uniform anticyclonic oscillation, visible by the phase shift of $\pi/2$ between the mean velocity components  $\langle U_x \rangle$ and $\langle U_y \rangle$. Its amplitude is of the order of $U_\mathrm{IW} \simeq 10^{-2} V_g$ (resp. $5 \times 10^{-4} V_g$) at the beginning (resp. end) of the decay. The gyration radius associated to this wave is $U_\mathrm{IW}  T_\mathrm{IW} / 2 \pi \simeq 3$~cm (resp. 1~mm).

In the non-rotating case, the mean horizontal shear persists over large times. Although very weak, it acts as a source of turbulence, which is found to affect the energy decay at large times (Sec.~\ref{sec:decay}). On the other hand, when rotation is present, this mean shear oscillates at a frequency $2 \Omega$ which is about 100 times larger than $\partial \langle U_x \rangle / \partial z$, yielding a strong suppression of the turbulence production. A nice consequence of this effect for the present experiment is that translating a grid in a rotating frame produces a turbulence which is much closer to homogeneity than in a non-rotating frame.

Finally, we note that the disturbance height associated to both the GW and IW flows, estimated as $\xi = (h / g)^{1/2} U$ where $U$ is the wave velocity, is less than 0.6~mm, yielding negligible surface slopes. As a consequence, contrarily to the free surface disturbances  of small wavelength discussed in \S~\ref{sec:fsd}, no significant optical distortion is expected here from those large scale waves.

\subsection{Large scale flows subtraction}
\label{sec:ave}

To summarize, the measured velocity field ${\bf U}$ can be written as the sum of 5 contributions: the 3 large scale flows (LSC, GW and IW), the 
turbulent field of interest ${\bf u}$, and the apparent
velocity field induced by the free surface disturbances
${\bf u}_\mathrm{FSD}$ (present only at early time for the measurements in the horizontal plane). The three large scale flows are simply described by their amplitude, $U_\mathrm{LSC}$, $U_\mathrm{GW}$ and $U_\mathrm{IW}$, which are slowly decaying in time. Keeping only the dominant spatial dependences of those 5 contributions, the measured velocity can be approximated as
\begin{eqnarray}
{\bf U}^{(n)}(x,y,z,t) & \simeq & U_\mathrm{LSC}(z,t) {\bf e}_x + U_\mathrm{GW}(t) \cos \left( \frac{2 \pi t}{T_\mathrm{GW}} \right) {\bf e}_x \nonumber \\
 & & + U_\mathrm{IW}(z,t) \left[ \cos \left(\frac{2 \pi t}{T_\mathrm{IW}} \right) {\bf e}_x + \sin \left(\frac{2 \pi t}{T_\mathrm{IW}} \right) {\bf e}_y \right] \nonumber \\
 & & + {\bf u}^{(n)}(x,y,z,t) + {\bf u}^{(n)}_\mathrm{FSD} (x,y,z,t),
\label{eq:sumu}
\end{eqnarray}
where $n$ is the realization number (the phase origin of the GW and IW flows are not considered for simplicity). Within this approximation, the three large scale flows  are uniform translations in the horizontal plane. As a consequence, providing that the turbulent scale is significantly smaller than the imaged size, these large scale flows can be readily subtracted from the measured velocity fields.  For the measurements in the horizontal plane, neglecting ${\bf u}_\mathrm{FSD}$, one has:
\begin{equation}
u_\alpha^{(n)} (x,y,t) = U_\alpha^{(n)} (x,y, t) - \langle U_\alpha^{(n)} (x,y,t) \rangle_{x,y,e},
\label{eq:subu}
\end{equation}
with $\alpha = x,y$. Here the brackets $\langle \cdot \rangle_{x,y,e}$ denote both ensemble and spatial averages. For a field $A^{(n)}({\bf r}, t)$ the ensemble average $\langle \cdot \rangle_e$  is defined as
$$
\langle A^{(n)}({\bf r}, t) \rangle_e = \frac{1}{N_r} \sum_{n=1}^{N_r}  A^{(n)}({\bf r}, t),
$$
where $n=1..N_r$ is the realisation ($N_r=4$ and 6 for the measurements in the vertical and horizontal plane respectively). The spatial average along the direction $x_\alpha$ is defined as
$$
\langle A^{(n)}({\bf r}, t) \rangle_{x_\alpha} = \frac{1}{L_\alpha} \int_0^{L_\alpha}  A^{(n)}({\bf r}, t) dx_\alpha,
$$
where $L_\alpha$ is the image size in the ${\bf e}_\alpha$ direction. Similarly, for the measurements in the vertical plane, the
average $\langle \cdot \rangle_{x,z,e}$ is computed. In order to reduce the statistical noise owing to the limited number of realisations, a temporal smoothing is also performed, on a range $[t_1, t_2]$ such that $t_2 - t_1 < (t_1 + t_2) / 20$. When there is no ambiguity, single brackets $\langle \cdot \rangle$ denotes in the following both the ensemble and spatial averages and the temporal smoothing. Finally, the root mean square (rms) is noted $A' = \langle A^2 \rangle^{1/2}$.

It must be noted that, at the end of the decay, when the
scale of motion becomes of the order or even larger than the imaged size,
distinguishing the large scale and the turbulent contributions becomes
ambiguous, and the subtraction may underestimate the actual
turbulent energy.

\section{Energy and integral scales}
\label{sec:decay}

\subsection{Energy decay}

\begin{figure}
\begin{center}
\includegraphics[width=6.5cm]{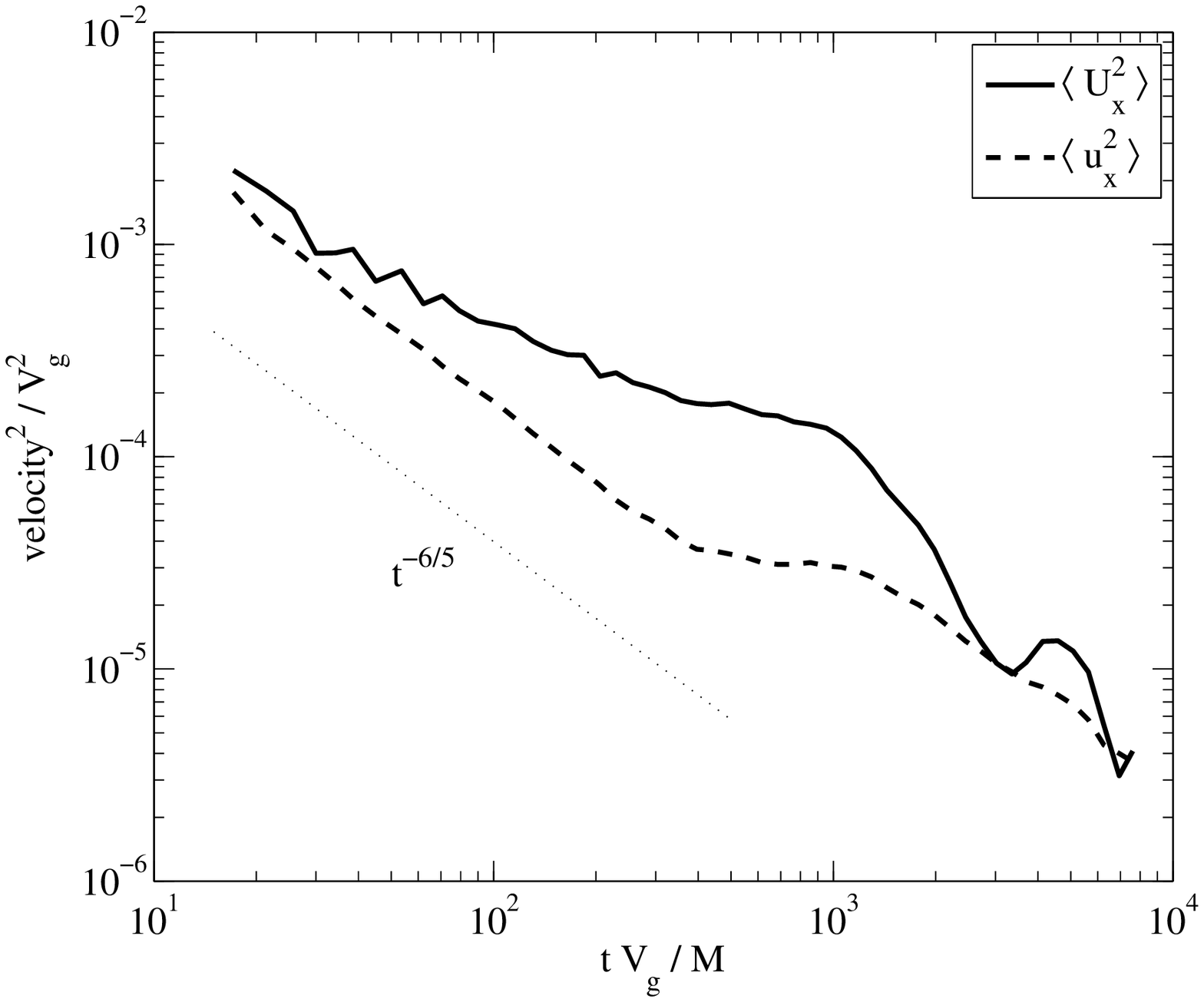}
\includegraphics[width=6.5cm]{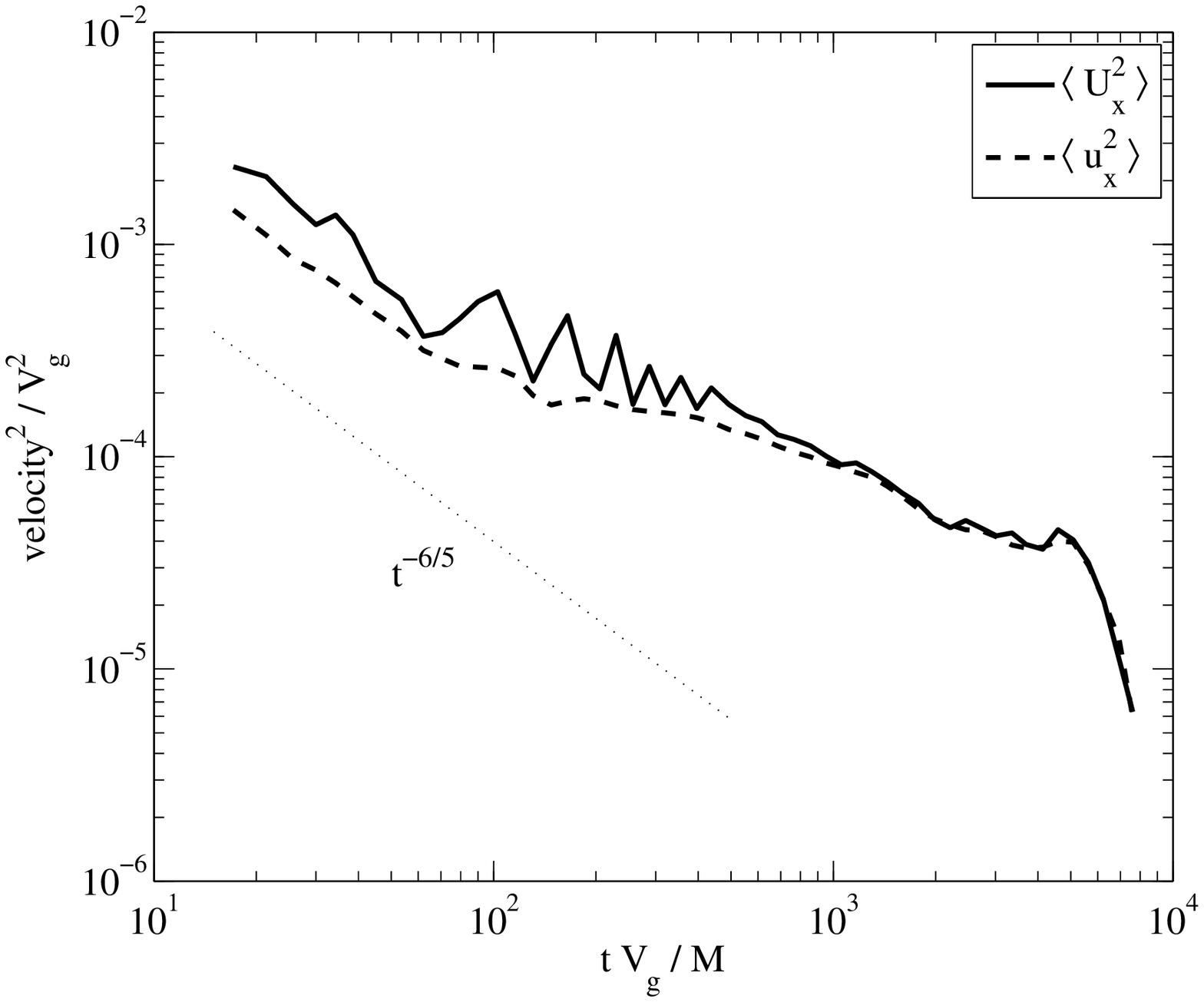}
\caption{Total and turbulent kinetic energy (streamwise variance). (a), $\Omega = 0$. (b), $\Omega = 0.10$~rad~s$^{-1}$ ($T=60$~s). The oscillations in the rotating case correspond to IW flow, of period $T_\mathrm{IW} = T/2 = 30$~s.} 
\label{fig:ken0}  
\end{center}                                            
\end{figure}                                             

The time evolution of the streamwise velocity variance for the total flow, $\langle U^2_x \rangle$, and the turbulent flow, $\langle u^2_x \rangle = \langle (U_x - \langle U_x \rangle)^2 \rangle$, are shown in Figure~\ref{fig:ken0}. In the absence of rotation (figure~\ref{fig:ken0}a), the energy of the LSC flow clearly dominates the total energy, by a factor up to 10 for $t \simeq 1000 M/V_g$.  On the other hand, when rotation is present (figure~\ref{fig:ken0}b), the turbulent energy is very close to the total energy, confirming that the mean flow is significantly reduced in the presence of rotation.

\begin{figure}
\begin{center}
\includegraphics[width=6.5cm]{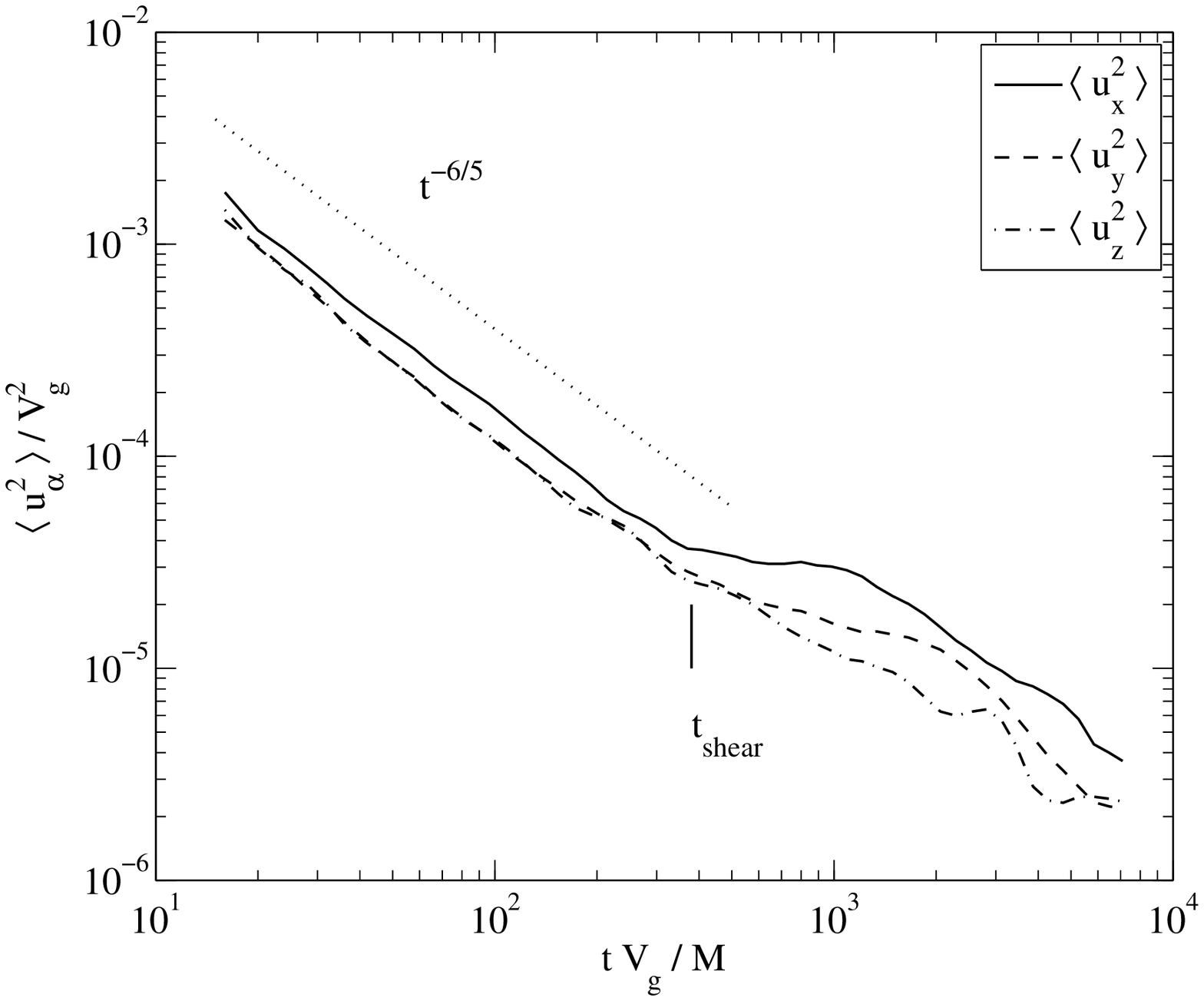}
\includegraphics[width=6.5cm]{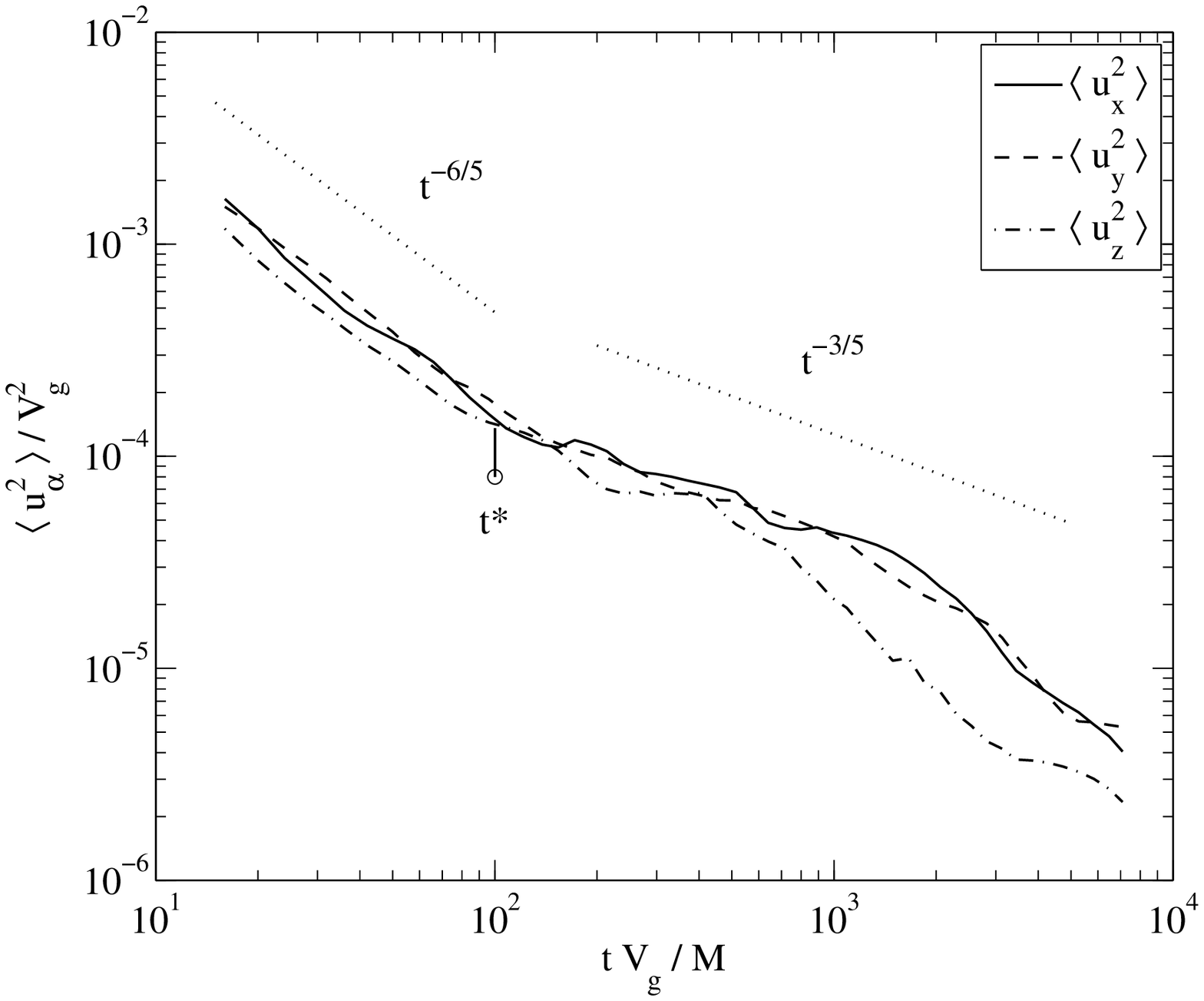}
\caption{Time evolution of the variance of the 3 velocity components. The variances $\langle u_x^2 \rangle$ and $\langle u_y^2 \rangle$ are computed from the horizontal PIV fields (camera C1), and $\langle u_z^2 \rangle$ from the vertical PIV fields (camera C2), for non-simultaneous experiments. (a), $\Omega = 0$. The vertical tick $t_{\rm shear}$ indicates the time after which the turbulent energy production by the residual mean shear becomes significant. (b), $\Omega = 0.05$~rad~s$^{-1}$. The vertical tick $t^*$ indicates the transition between the $t^{-6/5}$ isotropic decay and the $t^{-3/5}$ decay affected by the rotation.} 
\label{fig:varxyz}  
\end{center}                                            
\end{figure}

Once the mean flow is subtracted, the turbulent energy in the non-rotating case
decays as $t^{-n}$ up to $t \simeq 400 M/V_g$, with $n \simeq 1.22 \pm 0.05$ (figures~\ref{fig:ken0} and \ref{fig:varxyz}a). This decay exponent is very close to the Saffman (1967) prediction $n=6/5$ for unbounded turbulence. The streamwise variance, $\langle u_x^2 \rangle$, is approximately 1.4 times larger than the 2 spanwise variances $\langle u_y^2 \rangle$ and $\langle u_z^2 \rangle$, reflecting the usual residual anisotropy of grid turbulence (Comte-Bellot \& Corrsin, 1965). For $t > 400 M/V_g$, however, the shallower decay originates from the turbulence production by the mean vertical shear, which is a specific feature of the non-rotating case. This transition time, noted $t_{\rm shear}$ in figure~\ref{fig:varxyz}(a), is indeed of the order of the shear timescale, $(\partial U_x / \partial z)^{-1} \simeq 250$~s~$ \simeq 540 M/V_g$. The ordering of the 3 velocity variances for $t \gg t_{\rm shear}$, $\langle u_x^2 \rangle > \langle u_y^2 \rangle > \langle u_z^2 \rangle$, actually confirms the shear-dominated nature of the turbulence in the non-rotating case at large times (Tavoularis \& Karnik, 1989).

In non-dimensional form, the decay law of the streamwise variance  (neglecting possible time origin shift) writes
\begin{equation}
\frac{\langle u_x^2 \rangle}{V^2_g} \simeq A \left( \frac{t V_g}{M} \right)^{-6/5}.
\label{eq:dec0}
\end{equation}
A best fit yields a decay coefficient $A \simeq 0.045 \pm 0.005$ for $t < t_{\rm shear}$, a value in good agreement with the literature for grid turbulence (see, e.g., the review by Mohamed and LaRue, 1990). This low value reflects the weak efficiency of a grid to produce turbulent fluctuations.  This demonstrates that, in spite of the significant large scale flow generated by the present forcing, the decay of the turbulent kinetic energy  is close to that of classical grid turbulence, suggesting a negligible coupling between the mean flow and the small scale turbulence, at least for $t < t_{\rm shear}$.

The time evolution of the 3 velocity variances in the rotating case are shown in figure~\ref{fig:varxyz}(b) for $\Omega = 0.05$~rad~s$^{-1}$.  At early time, the 3 curves are very close to the reference case $\Omega = 0$ (figure~\ref{fig:varxyz}a), confirming that the rotation has no measurable effect at large Rossby numbers. After a crossover time $t^* \simeq 100 M/V_g$, the decay of the 2 horizontal variances  $\langle u_x^2 \rangle$ and  $\langle u_y^2 \rangle$ become shallower, showing a clear reduction of the energy decay by the rotation. On the other hand, the vertical variance $\langle u_z^2 \rangle$ first follows the horizontal variance short after the crossover time $t^*$, but sharply decreases soon after, reflecting a growth of anisotropy.  Since here the turbulence production by the mean shear is essentially suppressed by the background rotation, this departure from the $t^{-6/5}$ decay and the resulting anisotropy growth can now be interpreted as a pure effect of the rotation.  The anisotropy of the flow will be characterized in more details in Sec.~\ref{sec:aniso}, and we focus here on the behaviour of the horizontal variances only. 

\subsection{Crossover between the two decay regimes}

\begin{figure}
\begin{center}
\includegraphics[width=9.5cm]{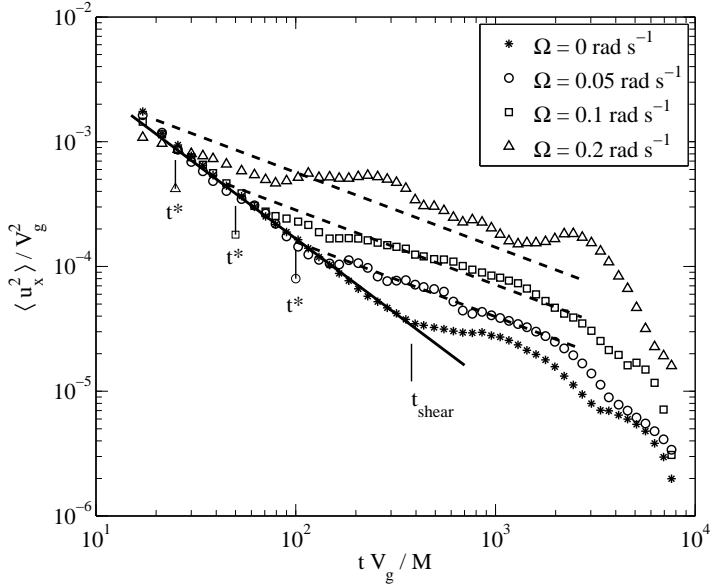}
\caption{Time evolution of the streamwise velocity variance $\langle u_x^2 \rangle$, for the non-rotating and the 3 rotating experiments. The solid line shows $A (t V_g / M)^{-6/5}$ and the dashed lines show $A_\Omega Ro_g^{-3/5} (t V_g / M)^{-3/5}$. The transition between the non-rotating ($t^{-6/5}$) and rotating ($t^{-3/5}$) decay laws occurs at $t^*$, indicated by the 3 vertical ticks for each rotation rate. For $t>t_{\rm shear}$ the turbulent energy production by the residual mean shear becomes significant in the non-rotating case.} 
\label{fig:ken}  
\end{center}                                            
\end{figure}

In order to characterize the influence of the rotation on the transition time $t^*$, the decays of the streamwise velocity variance $\langle u_x^2 \rangle$ are compared in figure~\ref{fig:ken} for the 4 sets of experiments. The crossover time $t^*$ decreases from 100 to approximately $30 M/V_g$ as $\Omega$ is increased, corresponding to approximately 0.4 tank rotation.  The small value of $t^*$ found for the highest rotation rate ($\Omega = 0.20$~rad~s$^{-1}$) indicates that the turbulent energy production in the wake of the grid may be indeed already affected by the background rotation in this specific case (the grid Rossby number is $Ro_g = 5.1$ only). The energy decay for this high $\Omega$ is indeed particular, showing unexpected large fluctuations of the streamwise velocity variance.

In the limit of large rotation rate, the energy decay can be modeled by assuming that the energy transfer rate scales as the linear timescale $\Omega^{-1}$. Based on this argument, Squires \etal (1994) proposed, using dimensional analysis, the following asymptotic decay law,
\begin{equation}
\frac{\langle u_x^2 \rangle}{V^2_g} \simeq A_\Omega Ro_g^{-3/5} \left( \frac{t V_g}{M} \right)^{-3/5},
\label{eq:decw}
\end{equation}
with $A_\Omega$ a non-dimensional constant.  Although the elapsed time is moderate here, the decay curves in figure~\ref{fig:ken} are actually compatible
with this shallower decay (\ref{eq:decw}).  Fitting the data for $\Omega = 0.05$ and $0.10$~rad~s$^{-1}$ yields $A_\Omega \simeq 0.020 \pm 0.005$ (the data at $\Omega = 0.20$~rad~s$^{-1}$ being excluded for the reason given before). Accordingly, the crossover time $t^*$ between the non-rotating and the rotating decay laws is obtained by equating Eqs.~(\ref{eq:dec0}) and (\ref{eq:decw}),
\begin{equation}
\frac{t^* V_g}{M} \simeq \left(\frac{A}{A_\Omega}\right)^{5/3} Ro_g \simeq (5 \pm 1) Ro_g,
\label{eq:tstar}
\end{equation}
yielding the values 100, 50 and 25 for the 3 rotation rates, which reproduces correctly the observed $t^*$ in figure~\ref{fig:ken}, at least for $\Omega = 0.05$ and $0.10$~rad~s$^{-1}$ (see the vertical ticks). Expressing this crossover time (\ref{eq:tstar}) in terms of the rotation rate is consistent with a transition occurring at fixed fraction of tank rotation,
$$
\frac{\Omega t^*}{2\pi} \simeq (5 \pm 1) / 4\pi \simeq 0.4 \pm 0.1.
$$
This result indicates that, after only half a tank rotation, the turbulence decay is dominated by the rotation.

It is remarkable that the transition between the two regimes $t^{-6/5}$ and $t^{-3/5}$ is sufficiently sharp, so that the analysis of Squires \etal (1994) can be recovered to a correct degree of accuracy. A similar transition in the form $t^{-10/7} \rightarrow t^{-5/7}$, with again a factor 2 between the non-rotating and the rotating decay exponents, has been observed in the recent simulation of Bokhaven \etal (2008), the discrepancy with the present exponents being probably associated to different energy content at small wavenumber in the experiment and the simulation. The steeper decay laws reported by Morize \& Moisy (2006a), with a transition from $t^{-2}$ to $t^{-1}$ as $\Omega$ is increased, is probably an effect of the significant lateral confinement present in that experiment.

\subsection{Integral scales}

The integral scales in the horizontal plane are useful to define the instantaneous turbulent Reynolds and Rossby numbers. They are defined as
\begin{equation}
L_{\alpha \alpha, \beta}(t) = \int_0^{\infty}  C_{\alpha \alpha,
\beta} (r,t) \, dr,
\label{eq:Lth}
\end{equation}
from the two-point correlation function of the $\alpha$ velocity component along the $\beta$ direction,
\begin{equation}
C_{\alpha \alpha, \beta} (r,t) =
\frac{\langle u_\alpha({\bf x}, t) u_\alpha({\bf x} + r {\bf
e}_\beta, t) \rangle}{\langle u_\alpha^2 \rangle}.
\label{eq:Cth}
\end{equation}
Computing $L_{\alpha \alpha, \beta}$ from experimental data using directly Eq.~(\ref{eq:Lth}) is not possible, because of the limited number of independent realizations for separations $r$ approaching the image size. To limit this uncertainty, the integral in Eq.~(\ref{eq:Lth}) is truncated at a suitably defined decorrelation length $r^*$,
\begin{equation}
L_{\alpha \alpha, \beta} = \int_0^{r^*}  C_{\alpha \alpha, \beta}
(r) \, dr, \quad \mbox{with  } C_{\alpha \alpha, \beta} (r^*) = C_{tr}.\label{eq:LCex}
\end{equation}
A threshold value of $C_{tr} = 0.2$ was found to give reliable results throughout the decay. At early time, when a reliable estimate of the full integral (\ref{eq:Lth}) can be achieved, the truncated estimate is found to underestimate the actual one by a factor of about 1.3 to 1.5. Although this definition systematically under-estimates the actual integral scales, the trends observed from these quantities are expected to represent the evolution of the true length scales.

It must be noted that the presence of large scale flows discussed in \S~\ref{sec:meanflow} would imply an unphysical increase of the horizontal correlations $C_{\alpha \alpha, \beta}(r)$, and consequently of the integral scales, if computed directly from the total measured velocity ${\bf U}$. Accordingly, the subtraction of the large scale flows is critical for reliable measurement of the integral scales.

\begin{figure}
\begin{center}
\includegraphics[width=8cm]{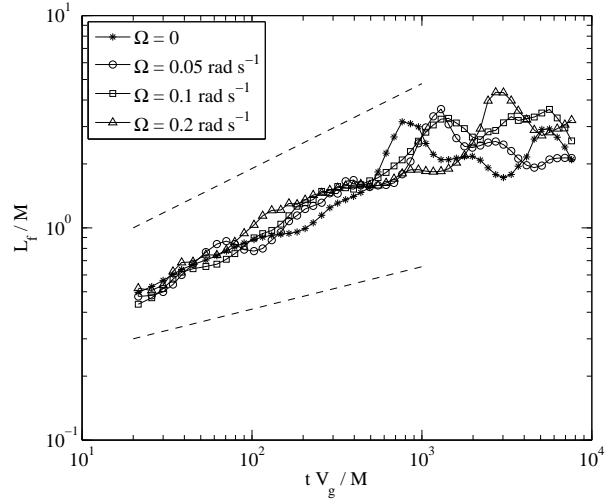}
\caption{Time evolution of the longitudinal integral scale in the horizontal plane, $L_f = (L_{11,1} + L_{22,2})/2$, for the 4 series of experiments. The upper and lower dashed lines show the scalings $t^{2/5}$ and $t^{1/5}$ for reference.} 
\label{fig:Lf}  
\end{center}                                            
\end{figure}                                             

The time evolution of the longitudinal integral scale, averaged over the two horizontal directions $x$ and $y$ (noted here 1 and 2 by convention),
$$
L_f = \frac{1}{2}(L_{11,1} + L_{22,2}),
$$
is plotted in figure~\ref{fig:Lf}, and shows little influence of the background rotation. A best fit yields a power law $L_f(t) \simeq t^{0.35 \pm 0.05}$, for $t < 1000 M/V_g$, for all rotation rates. The scatter at larger time is probably a consequence of the inadequate subtraction of the mean flow, which may occur when the size of the largest vortices becomes comparable to the imaged area.

Dimensional analysis actually predicts different growth laws for $L_f$ in the non-rotating and rotating cases (Squires \etal 1994),
\begin{eqnarray}
\frac{L_f}{M} \simeq B \left( \frac{t V_g}{M} \right)^{2/5} & \qquad (t \ll t^*), \nonumber \\
\frac{L_f}{M} \simeq B_\Omega Ro_g^{1/5} \left( \frac{t V_g}{M} \right)^{1/5} & \qquad (t \gg t^*),
\label{eq:lw}
\end{eqnarray}
with $B$ and $B_\Omega$ non-dimensional constants. Surprisingly, although the $t^{-6/5} \rightarrow t^{-3/5}$ transition at $t=t^*$ is evident in the energy decay curves (figure~\ref{fig:ken}), there is no evidence for the equivalent $t^{2/5} \rightarrow t^{1/5}$ transition for $L_f$ in figure~\ref{fig:Lf}. Within the experimental uncertainty, a single power law $t^{2/5}$ actually provides a reasonable description for the growth of $L_f$ both in the non-rotating and in the rotating cases.

\subsection{Instantaneous Reynolds and Rossby numbers}
\label{sec:rero}

\begin{figure}
\begin{center}
\includegraphics[width=6.5cm]{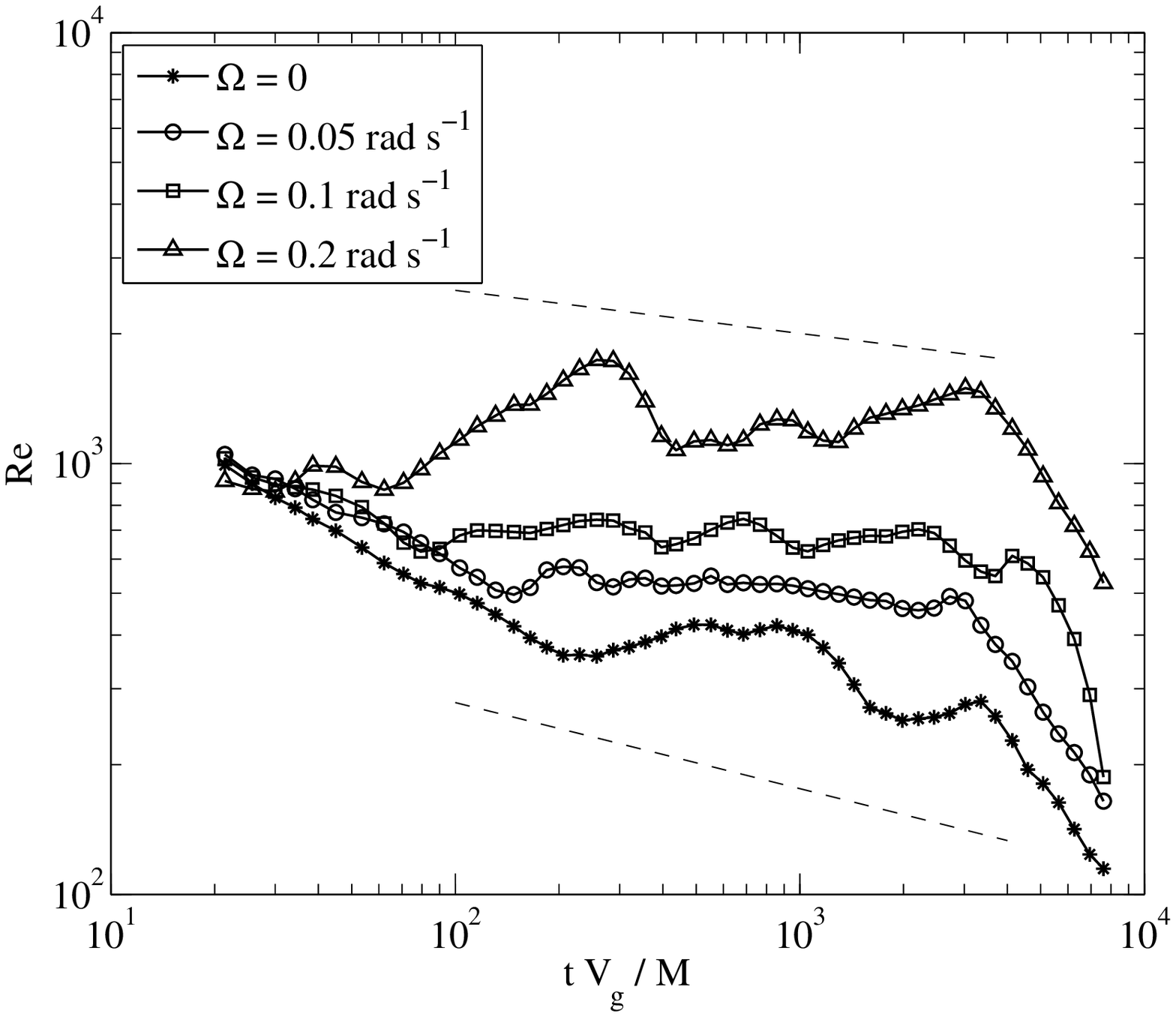}
\includegraphics[width=6.5cm]{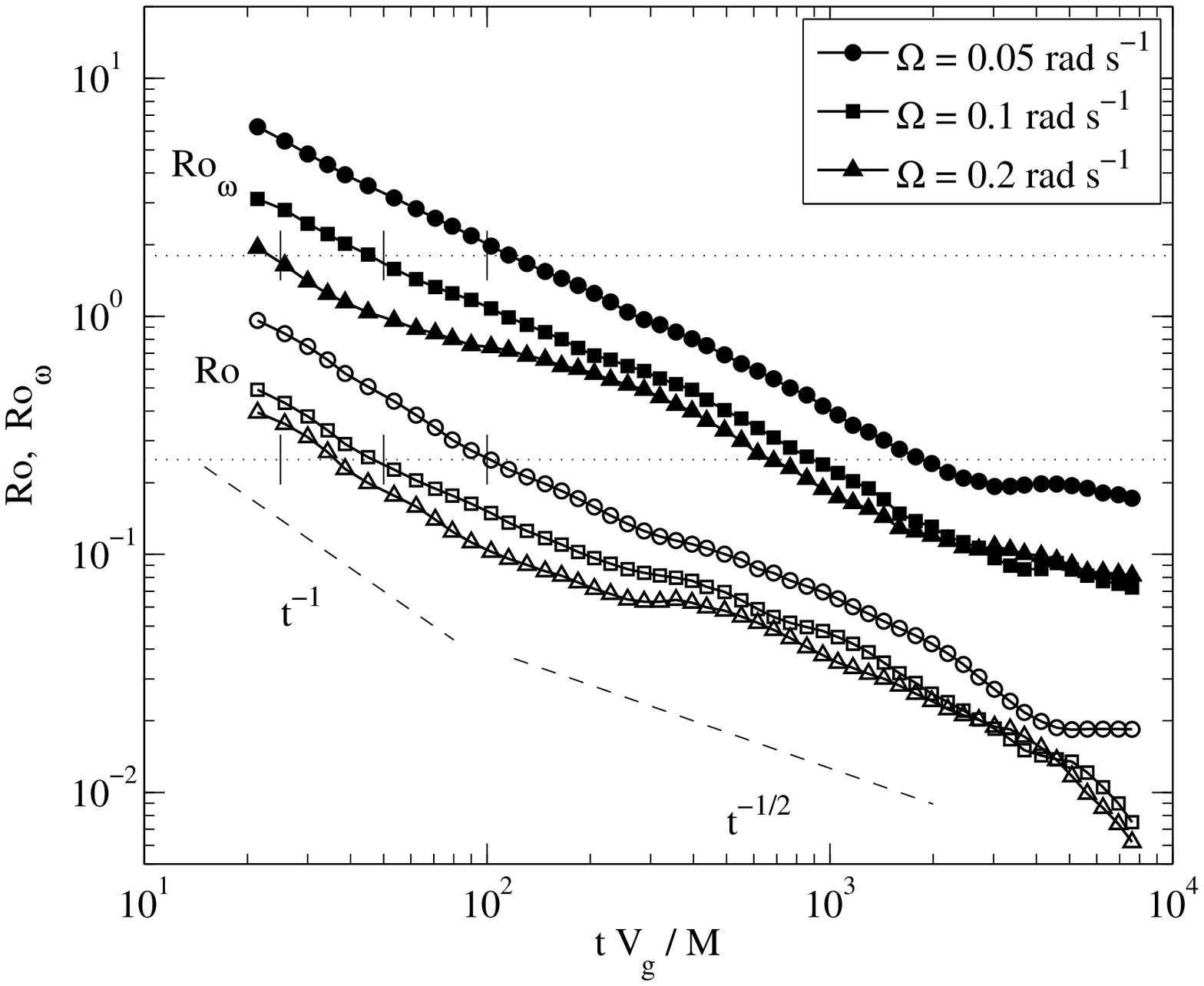}
\caption{(a), Reynolds number $Re(t) = u'_x L_f / \nu$. (b), Micro- and Macro- Rossby numbers, for the three experiments with background rotation. Upper curves (black symbols): $Ro_\omega = \omega_z' / 2\Omega$. Lower curves (open symbols): $Ro = u_x'/ 2 \Omega L_f$. The horizontal dotted lines show the thresholds, $Ro_\omega = 1.8$ and $Ro = 0.25$, with the corresponding transition times $t^*$ indicated by the vertical ticks [see Eq.~(\ref{eq:tstar})]. The dashed lines show the scalings $t^{-1}$ and $t^{-1/2}$ for reference.} 
\label{fig:ReRo}  
\end{center}                                            
\end{figure}

The instantaneous Reynolds number, and the macro and micro Rossby numbers (Jacquin \etal 1990), are finally defined as
\begin{equation}
Re (t) = \frac{u_x' \, L_f}{\nu}, \qquad Ro(t) = \frac{u_x'}{2 \Omega \, L_f}, \qquad Ro_\omega(t) = \frac{\omega_z'}{2 \Omega}.
\label{eq:ReRo}
\end{equation}
The time evolution of those numbers are plotted in figure~\ref{fig:ReRo}(a-b). After a short period of sharp decay similar to the non-rotating case,  the Reynolds number in the rotating cases show a very weak decay in the range $t^* < t  < 3000 M/V_g$. Typical Reynolds numbers range from 500 to 1300 as the rotation rate is increased.

In spite of the absence of clear transition in the growth law for the integral scale, the time evolution of $Re(t)$ compares correctly with the forms expected for the non-rotating and rotating cases -- see Eq.~(\ref{eq:dec0}), (\ref{eq:decw}) and (\ref{eq:lw}),
\begin{eqnarray}
Re(t) \propto Re_g \left( \frac{t V_g}{M} \right)^{-1/5} & \qquad (t \ll t^*), \nonumber \\
Re(t) \propto Re_g Ro_g^{-1/10} \left( \frac{t V_g}{M} \right)^{-1/10} & \qquad (t \gg t^*). \nonumber
\end{eqnarray}

The micro-Rossby number $Ro_\omega$, shown in figure~\ref{fig:ReRo}(b), takes values about 10 times larger than $Ro$ throughout the decay (note that $Ro_\omega$ may be underestimated at small times because of the limited PIV resolution).  This moderate ratio indicates that the range between the large scales dominated by the rotation and the small scales is indeed limited for the Reynolds number of the present experiments.

The joint decay and growth laws for the velocity and integral scale actually lead to a remarkably simple decay law for the macro Rossby number $Ro(t)$. Combining again  Eq.~(\ref{eq:dec0}), (\ref{eq:decw}) and (\ref{eq:lw}) shows that, for $t<t^*$, the non-linear time scale $\tau_{nl} = L_f(t) / u_x'(t)$ is simply proportional to the elapsed time $t$, with no dependence on the initial grid time scale $M/V_g$. Since the Rossby number is given by $Ro = (2 \Omega \tau_{nl})^{-1}$, it turns out to be only a function of the number of tank rotations,
\begin{eqnarray}
Ro(t) \propto Ro_g \left( \frac{t V_g}{M} \right)^{-1} \propto (2\Omega t)^{-1} & \qquad (t \ll t^*), \nonumber \\
Ro(t) \propto Ro_g^{1/2} \left( \frac{t V_g}{M} \right)^{-1/2} \propto (2\Omega t)^{-1/2} & \qquad (t \gg t^*). \nonumber
\end{eqnarray}
At the transition $t = t^*$, which is reached after a fixed number of rotations, the Rossby number is indeed found approximately constant, $Ro(t^*) \simeq 0.25$ (see the vertical ticks at $t^*$ in figure~\ref{fig:ReRo}b). This value is in correct agreement with the transitional Rossby numbers reported by Hopfinger \etal (1982) and Staplehurst \etal (2008).

No such simple law applies for the micro-Rossby number $Ro_\omega(t)$. For $t \ll t^*$, assuming again isotropic turbulence, the decay law of $Ro_\omega$ can be inferred from the relation between the vorticity rms, the velocity rms and the dissipation rate,
\begin{eqnarray}
\epsilon = - \frac{1}{2} \frac{\partial {\bf u'}^2}{\partial t} = - \frac{3}{2} \frac{\partial u_x'^2}{\partial t} = \nu \boldsymbol{\omega}'^2 = 3 \nu \omega'^2_z.
\label{eq:epsiso}
\end{eqnarray}
Combining the isotropic decay law (\ref{eq:dec0}) with Eq.~(\ref{eq:epsiso}) yields
$$
Ro_\omega(t) = \sqrt{\frac{3}{5}} A^{1/2} Re_g^{1/2} Ro_g \left( \frac{t V_g}{M} \right)^{-11/10} \qquad (t \ll t^*),
$$
with a scaling exponent very close to that of $Ro(t)$. Evaluating $Ro_\omega$ at the transition $t \simeq t^*$, using Eq.~(\ref{eq:tstar}), finally yields
$$
Ro_\omega(t^*) \simeq \sqrt{\frac{3}{5}} (5 \pm 1)^{-11/10} A^{1/2} Re_g^{1/2} Ro_g^{-1/10}.
$$
Accordingly, no strictly constant micro-Rossby number is expected at the transition, although the dependence with the rotation rate, as $\Omega^{1/10}$, is very weak ($\Omega$ is varied by a factor of 4 only in the present experiment). As shown by the vertical ticks in figure~\ref{fig:ReRo}(b),  $Ro_\omega$ takes values which actually turn out to be approximately constant at the transition, $Ro_\omega(t^*) \simeq 1.8$. Interestingly, this value is close to the empirical threshold reported by Morize \etal (2005), below which the energy spectrum and the velocity derivative skewness were found to depart from the classical Kolmogorov predictions (computation of the macro-Rossby number based on the integral scale were not possible in that study). Although the macro-Rossby number is probably a more relevant parameter to describe this transition, the similar Reynolds number of the two experiments explains the similar values of $Ro_\omega$ found at the transition.

\section{Dynamics of the anisotropy}
\label{sec:aniso}

\subsection{Visualisation of the vertical layers}

\begin{figure}
\begin{center}
\includegraphics[width=9.5cm]{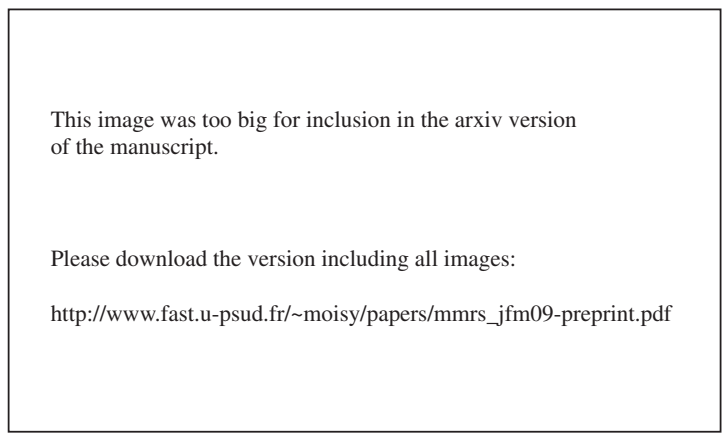}
\caption{Sequence of 6 snapshots of the velocity and spanwise vorticity $\omega_y$ in the vertical plane $(x,z)$ for $\W = 0.20$ rad.s$^{-1}$.  The imaged area is 1 m
$\times$ 1 m. The grid is translated from left to right, and the time origin $t=0$ is defined as the grid goes through the center of the imaged area. The color range is normalized by the rms $\omega_y' = \langle \omega_y^2 \rangle_{w,z,e}^{1/2}$ computed for each time.}
\label{fig:p30_wy}
\end{center}  
\end{figure}

We now focus on the growth of anisotropy in the vertical plane $(x,z)$. Figure~\ref{fig:p30_wy} shows a sequence of  6 snapshots of the velocity field and spanwise vorticity $\omega_y$ after the transition $t > t^*$, for $\W = 0.20$ rad~s$^{-1}$ (see also the Movie 3). The anisotropy can be visually detected from the first snapshot, and the presence of vertical layers of ascending or descending fluid becomes evident after 8 tank rotations (figure~\ref{fig:p30_wy}c). Although those layers are difficult to infer from the velocity field itself, because of the superimposed strong horizontal flow, they clearly appear through the surrounding layers of nearly constant $\omega_y$ of alternate sign. Those layers of vertical velocity are consistent with a trend towards a three-component two-dimensional flow, with zero vertical variations of the velocity field, $\partial {\bf u} / \partial z = 0$, but non-zero vertical velocity $u_z$ originating from the initial conditions.

As time proceeds, the vertical layers become thinner and more vertically coherent (note that since only the intersection of the layers with the measurement plane can be visualized, the apparent thickness may overestimate the actual one). At large time (figure~\ref{fig:p30_wy}e,f), although those layers are nearly coherent from the bottom wall up to the free surface, they are not strictly vertical, but rather show wavy disturbances. Those disturbances have amplitude and characteristic vertical size of the order of the layer thickness, suggesting the occurrence of a shear instability. We will examine in Sec.~\ref{sec:vor} the consequence of this instability on the dynamics and statistics of the vertical vorticity field.

\subsection{Decay of the vertical velocity and anisotropy growth}

\begin{figure}
\begin{center}
\includegraphics[width=6.5cm]{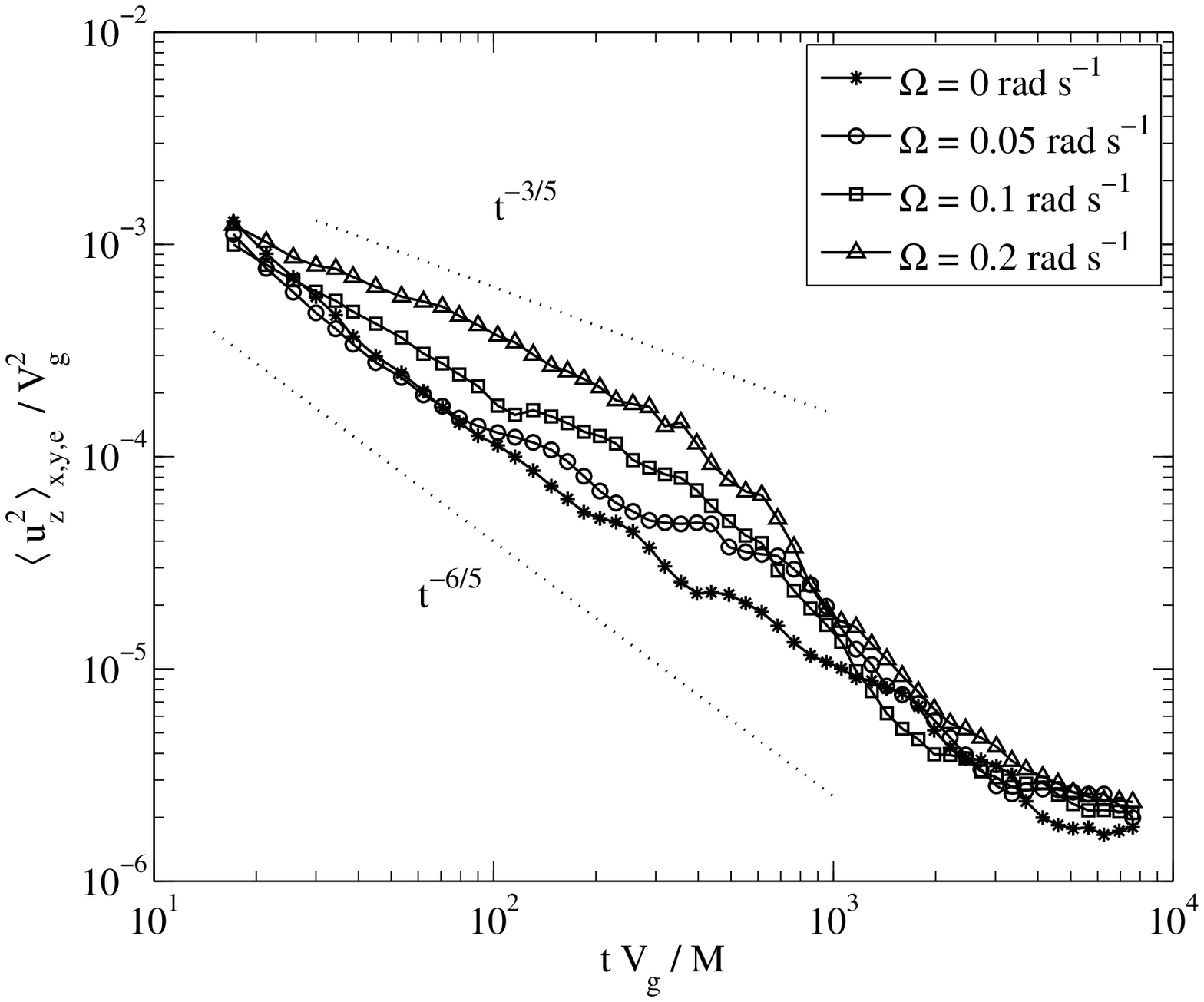}
\includegraphics[width=6.5cm]{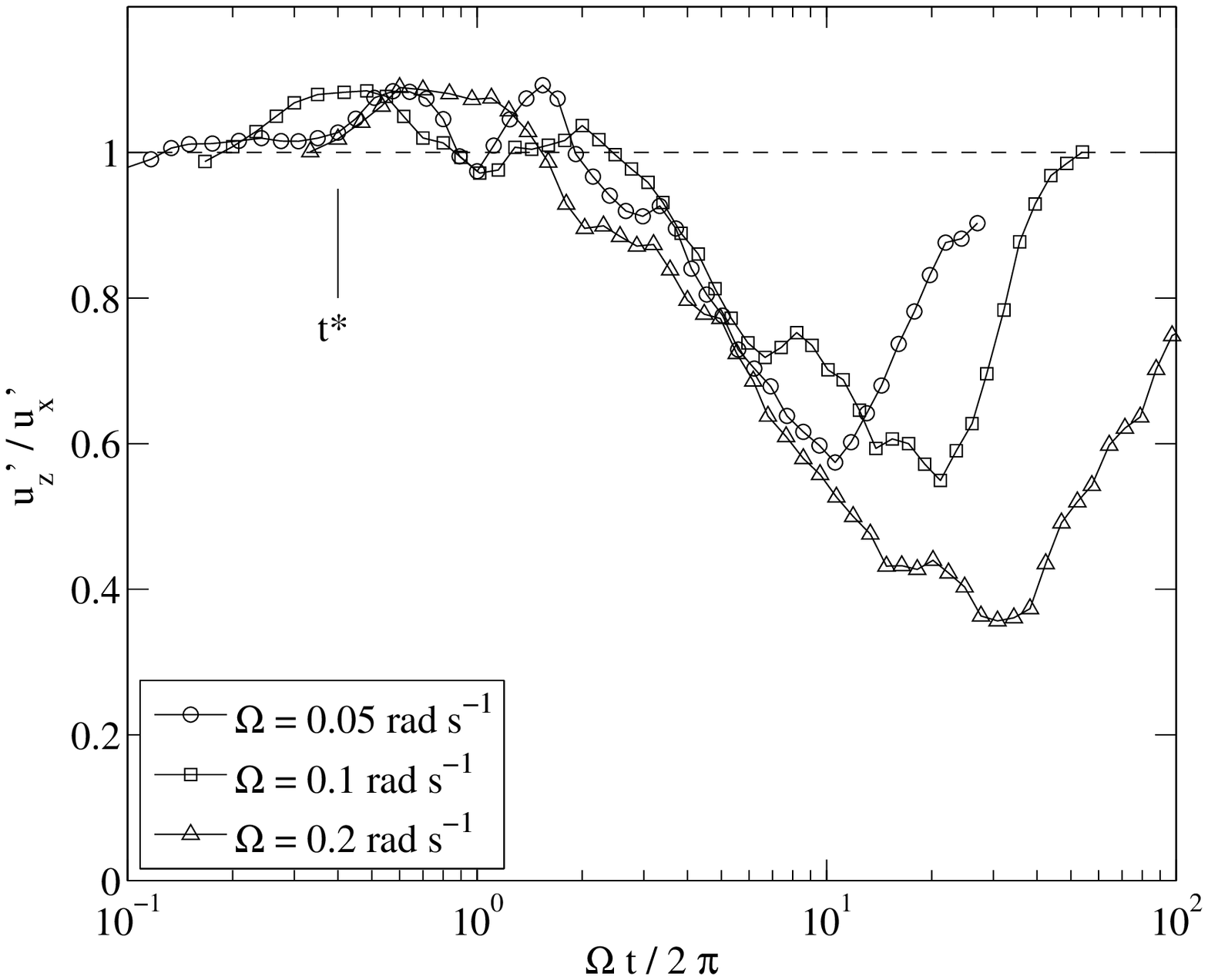}
\caption{(a), Time evolution of the vertical velocity variance $\langle u_z^2 \rangle$, for the non-rotating and the 3 rotating experiments. (b), Isotropy factor $u_z' / u_x'$ for the 3 rotating experiments, as a function of number of tank rotation. The vertical line indicates the transition between the $t^{-6/5}$ and the $t^{-3/5}$ decay regimes at $\Omega t^* / 2\pi \simeq 0.4$.}
\label{fig:iso}
\end{center}
\end{figure}
%

The time evolution of the vertical velocity variance $u_z'^2 = \langle u_z^2 \rangle_{x,z,e}$ and the isotropy ratio $u_z' / u_x'$,  plotted in figure~\ref{fig:iso}(a-b), show a complex and non-trivial behaviour. Here, the spatial average is computed only in the core of the flow, excluding layers of thickness $0.1 h$ near the bottom wall and the free surface.

Similarly to the horizontal variance (see figure~\ref{fig:ken}), the vertical variance for the rotating cases first departs from the reference curve $t^{-6/5}$ of the non-rotating case, and follows a shallower decay which is compatible again with a $t^{-3/5}$ law, at least in an intermediate temporal range. Although the $t^{-6/5} \rightarrow t^{-3/5}$ transition is not as sharp as for the horizontal variance, perhaps because of the limited statistics achieved for the measurements in the vertical plane, the transition time $t^*$ is compatible with the one determined for $\langle u_x^2 \rangle$, corresponding to $\Omega t^* / 2\pi \simeq 0.4$ tank rotation. Since both horizontal and vertical components follow the same decay law $t^{-3/5}$ short after $t^*$, the flow remains approximately isotropic, as shown by the constant ratio $u_z' / u_x' \simeq 1 \pm 0.1$ in figure~\ref{fig:iso}(b). However, this isotropy holds only during a restricted time range, until $\Omega t / 2 \pi \simeq 2$ tank rotations. After this time, corresponding to approximately $t \simeq (400-800) M/V_g$ in figure~\ref{fig:iso}(a), the vertical variance follows a significantly sharper decay,  whereas the horizontal variance still decays as $t^{-3/5}$, yielding a growing anisotropy consistent with the flow visualisations in the vertical plane.

Although the formation of vertical structures is evident in the spanwise vorticity field $\omega_y$ (figure~\ref{fig:p30_wy}), the anisotropy remains however moderate when expressed in terms of the velocity variances, as shown in figure~\ref{fig:iso}(b). The ratio $u_z' / u_x'$ reaches a weak minimum between 0.6 and 0.4 only, after $10-30$ tank rotations, depending on the rotation rate. Interestingly, the ratio $u_z' / u_x'$ for the different rotation rates collapse in the anisotropy growth regime when plotted as a function of the number of tank rotations $\Omega t / 2 \pi$, suggesting here again an essentially linear mechanism for the building of this anisotropy.

A remarkable feature of figure~\ref{fig:iso}(b) is the reverse trend $u_z' / u_x' \rightarrow 1$ observed at large time.  A similar behaviour is obtained for the Reynolds stress anisotropy in the numerical simulations of Morinishi, Nakabayashi \& Ren (2001). This apparent return to isotropy, which breaks the scaling with the linear timescale $\Omega^{-1}$, is associated to the flattening of the decay of $\langle u_z^2 \rangle$ at large time, visible in figure~\ref{fig:iso}(a). However,  the corresponding flow fields at large time (figures~\ref{fig:p30_wy}e,f) are clearly not isotropic, but rather show wavy thin vertical layers. 

\begin{figure}
\begin{center}
\includegraphics[width=6.5cm]{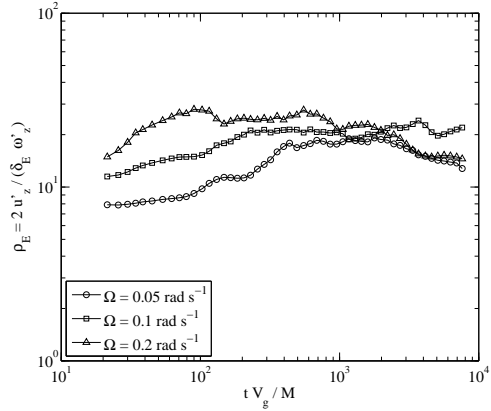}
\caption{
Time evolution of the Ekman ratio $\rho_E = 2 u'_z / \delta_E \omega'_z$.}
\label{fig:profuz}
\end{center}
\end{figure}

It must be noted that the magnitude of vertical velocity measured throughout the decays, even at large time,  remains comfortably larger than the one expected for the Ekman pumping induced by the horizontal flow.  According to the linear Ekman pumping theory, a quasi-2D field of vertical vorticity rms $\omega'_z$ would lead to a characteristic vertical velocity rms of ${u'}^{E}_z  = \delta_E \omega'_z / 2$ for $z \simeq \delta_E$ (Greenspan 1968), where $\delta_E$ is the Ekman layer thickness (see Sec.~\ref{sec:pivres}). The discrepancy between the actual $u_z'$ and the Ekman pumping estimate ${u'}^{E}_z$ may be therefore measured by the ratio
\begin{equation}
\rho_E = \frac{2 u'_z}{\delta_E \omega'_z}.
\label{eq:re}
\end{equation}
Note that this ratio corresponds, within a numerical prefactor, to the ratio between the Taylor scale $\lambda = \sqrt{15} u'_z / \omega'_z$ and $\delta_E$.
Figure~\ref{fig:profuz} shows the time evolution of $\rho_E$, computed from non-simultaneous measurements of $\omega_z'$ and $u_z'$ in the horizontal and vertical measurements respectively. Values around 10 at early time, slightly increasing up to about 20 at larger time, are obtained, confirming that effects of the Ekman pumping can be neglected in the present experiments. Accordingly, the vertical fluctuations found at late time originate from the initial vertical fluctuations induced by the grid.

It is remarkable that the time at which the ratio $\rho_E$ becomes constant corresponds approximately to the time $t^*$ at which the energy decay curves depart from the non-rotating reference curve. The initial increase of $\rho_E$ may be simply described from the decay laws of Sec.~\ref{sec:decay}.  From Eqs.~(\ref{eq:dec0}) and (\ref{eq:epsiso}), it follows that $\rho_E (t) \propto (2 \Omega t)^{1/2}$ for $t < t^*$,  in qualitative agreement with the data of figure~\ref{fig:profuz} at early time.  At the transition time $\Omega t^* / 2\pi \simeq 0.4$, the value of $\rho_E$ is therefore independent of $Ro_g$.  The fact that $u_z'$ and $\omega_z'$ remains approximately proportional for $t>t^*$ is an indication that the dynamics of the vertical velocity follows, in average, the dynamics of the vorticity. This supports the picture that the vertical velocity behaves as a scalar field, passively advected by the horizontal flow at large time.

\subsection{Integral scales in the vertical plane}

In order to relate the evolution of the vertical velocity variance to the formation, thinning and instability of the vertical layers, we now focus on the statistical geometry of those layers. For this, we have computed the 3 integral scales $L_{11,3}$, $L_{33,1}$ and $L_{33,3}$ from the velocity fields in the vertical plane, using definitions~(\ref{eq:Cth})-(\ref{eq:LCex}) with $\alpha, \beta = 1,3$.  $L_{11,3}$ characterizes the trends towards two-dimensionality, $L_{33,3}$ the vertical coherence of the layers, and $L_{33,1}$ the thickness of the layers.
No reliable measurement of $L_{11,1}$ could be obtained from the vertical fields, because of the ambiguity of the subtraction of the horizontal LSC flow at large time: Large scale vortices having their axis out of the measurement plane produce strong horizontal velocity which, if subtracted, yield an unphysical decrease of $L_{11,1}$. Here again, in Eq.~(\ref{eq:Cth}), the depth-average excludes lower and upper layers over a thickness of $h/10$, in order to avoid boundary effects. In the extreme case of an unbounded $z$-invariant 2D flow, the vertical correlations would be $C_{\alpha \alpha,3}(r)=1$, yielding $L_{\alpha \alpha,3}= \infty$. By convention, if $C(r)$ does not decrease below the threshold $C_{tr} = 0.2$ (see Eq.~(\ref{eq:LCex})), $L_{\alpha \alpha , \beta}$ is taken equal to the channel depth $h$.

\begin{figure}
\begin{center}
\includegraphics[width=8cm]{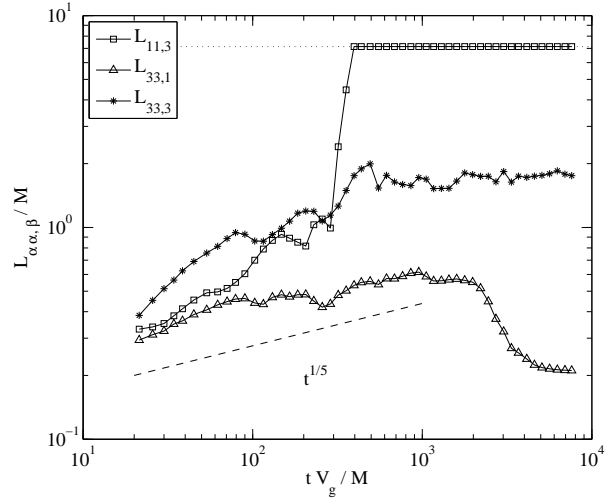}
\caption{Time evolution of the normalized integral scales
$L_{\alpha \alpha , \beta}/M$ computed in the vertical plane, for $\Omega = 0.10$~rad~s$^{-1}$. By convention, $L_{11,3}$ is taken equal the normalized tank height, $h/M \simeq 7$ (upper dotted line), when the vertical correlation of the horizontal velocity, $C_{11,3}$, does not decrease below 0.2.} 
\label{fig:Laab}
\end{center}
\end{figure}

The typical time evolution of the 3 integral scales is shown in figure~\ref{fig:Laab}, in the case $\Omega = 0.10$~rad~s$^{-1}$. At short time, the longitudinal integral scale $L_{33,3}$ is, as expected, larger than the two transverse ones (one has $L_{33,3} = 2 L_{11,3} = 2 L_{33,1}$ for isotropic turbulence). The most spectacular effect is the rapid growth of $L_{11,3}$, characterising the vertical correlation of the horizontal velocity, which is a clear signature of the two-dimensionalisation of the large scales of the flow. This rapid growth leads to a saturation of the integral scale for $t V_g / M > 400$. Although the data are too noisy to check the scaling of this saturation time with the rotation rate, it occurs roughly at a constant number of tank rotations as $\Omega$ is varied, which again supports the idea of a linear mechanism: At early time, energy is contained at scale $L_{11, 1} \simeq M$, and at each tank rotation eddies grow vertically by wave propagation,  so $L_{11, 3}$ increases by an amount of $L_{11, 1}$ (Davidson \etal 2006). Accordingly, $L_{11, 3}$ is expected to reach the channel depth $h$ after a number of tank rotations of order of $h/M \simeq 7$, which is in qualitative agreement with the present observations.

After the saturation of $L_{11,3}$, the vertical correlation of the vertical velocity remains constant until the end of the experiment, with $L_{33,3} \simeq 2 M$, indicating a significant, although finite, vertical coherence of the ascending and descending plane jets. Note however that even strictly coherent thin layers of constant velocity would lead to finite integral scale $L_{33,3}$, because the tilting of the layers by the oscillating shear of the IW flow strongly reduces the vertical correlation as the layers become thinner.


A remarkable feature of figure~\ref{fig:Laab} is the sharp decrease of the horizontal correlation of the vertical velocity, described by $L_{33,1}$, for $t V_g / M > 2000$ (corresponding to $\Omega t / 2\pi \simeq 16-24$ tank rotations), and its subsequent saturation to the very low value $L_{33,1} \simeq 0.2 M \simeq 30$~mm at large time. As far as the integral scales are concerned,  the turbulence becomes highly anisotropic in the final stage of the decay, showing a non-trivial ordering (see figures~\ref{fig:Lf} and \ref{fig:Laab})~:
\begin{equation}
L_{33,1} \ll L_{33,3} \simeq L_{11,1} \ll L_{11,3}.
\label{eq:order}
\end{equation}
The low asymptotic value of $L_{33,1}$ suggests that, in the final regime, the vertical velocity fluctuations have a well defined characteristic scale in the horizontal direction, i.e. there is no global vertical motion at scales larger than the thickness of the layers.  This final value of $L_{33,1}$, which provides an estimate for the average thickness of the layers, is found to slightly decrease, from $0.23 M$ to $0.17 M$, as $\Omega$ is increased, suggesting that the thinning of the layers induced by the horizontal straining motion due to the large scale vortices is stronger at high rotation rate.

\subsection{Thinning and instabilities of the vertical layers}
\label{sec:rti}

\begin{figure}
\begin{center}
\includegraphics[width=6.5cm]{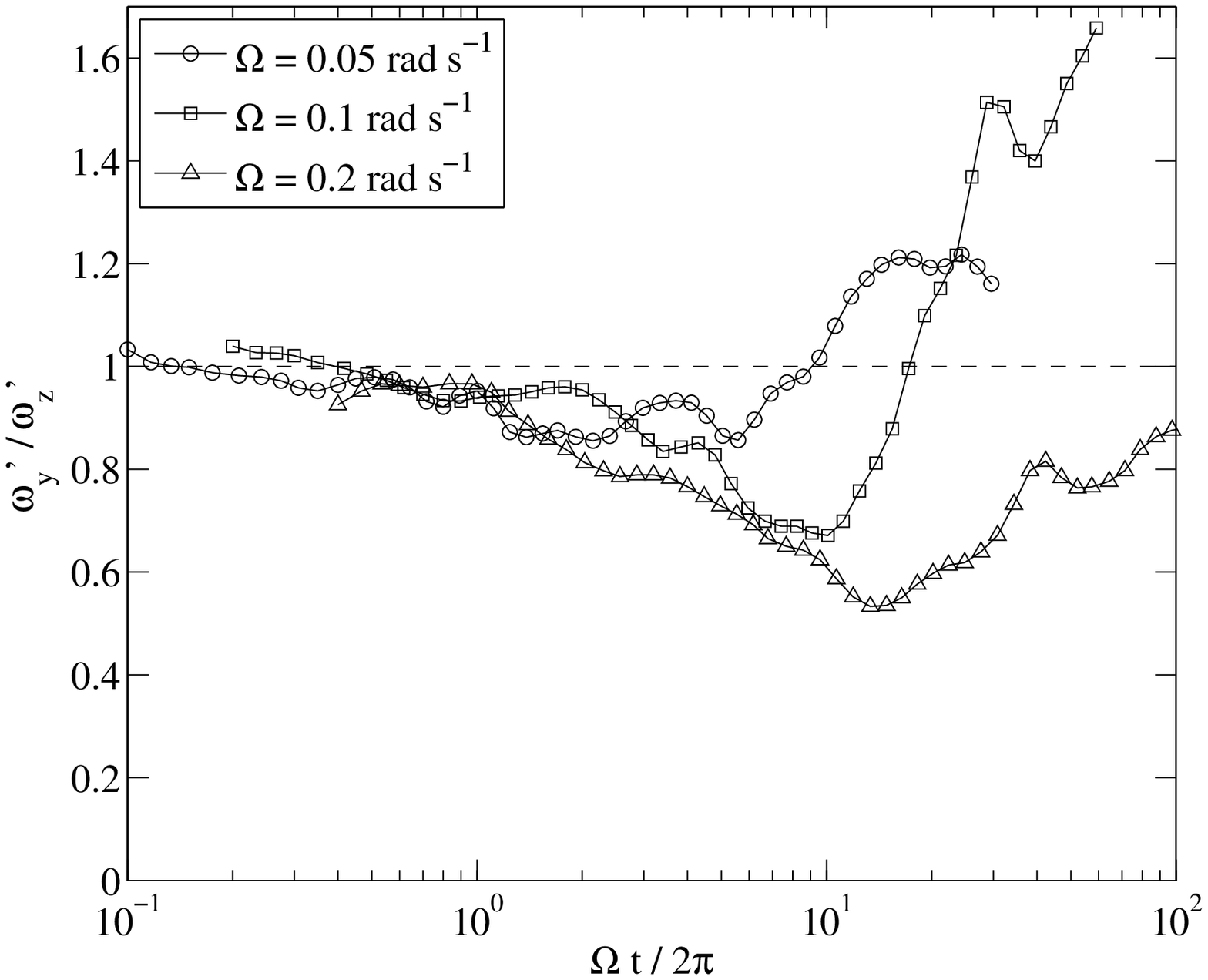}
\includegraphics[width=6.5cm]{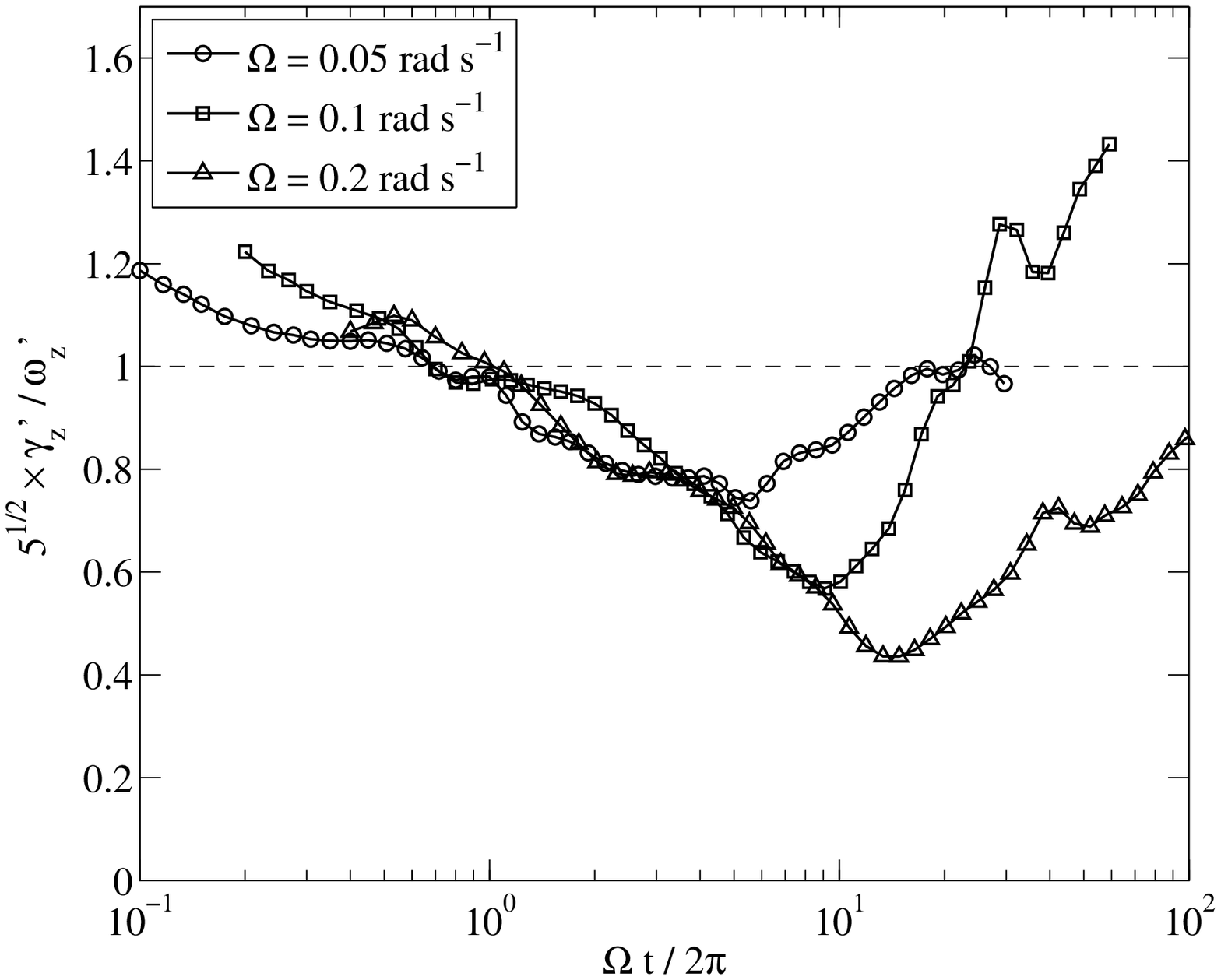}
\caption{Time evolution of the velocity gradient isotropy factors: (a), $\omega'_y / \omega'_z$; (b), $\sqrt{5} \gamma'_z / \omega'_z$. In each figure, the horizontal dashed line indicates the isotropic values, $\omega'_y / \omega'_z= \sqrt{5} \gamma'_z / \omega'_z = 1$.} 
\label{fig:divrot}  
\end{center}                                            
\end{figure}                                             

In spite of the strong anisotropy of the integral scales in the final stage (\ref{eq:order}), the velocity variances showed an unexpected return to isotropy, $u_z' / u_x' \rightarrow 1$ (figure~\ref{fig:iso}b). This apparent return to isotropy is also present at small scales, as shown by the following two velocity gradient isotropy factors,
\begin{equation}
\frac{\omega_y'}{\omega_z'} \qquad \mbox{and} \qquad \sqrt{5} \frac{\gamma_z'}{\omega_z'},
\label{eq:isofac}
\end{equation}
where $\gamma_z' = \langle (\partial u_z / \partial z)^2 \rangle^{1/2}$ is the rms of the vertical strain rate. The vertical strain rate plays an important role, as it is responsible for the stretching of the absolute vertical vorticity. In isotropic turbulence, both quantities are equal to 1 (the second equality follows from the classical isotropic relation $\epsilon = 15 \nu \gamma_z'^2 = \nu \boldsymbol{\omega}'^2 = 3 \nu \omega_z'^2$, where $\epsilon$ is the dissipation rate). For a two-dimensional flow with arbitrary vertical velocity, one has $\gamma_z = 0$, whereas $\omega_y = 0$ is true only for a two-dimensional two-component flow. As a consequence, the two isotropy factors may be considered as signatures of the {\it dimensionality} and {\it componentality} of the small scales respectively.

Figure~\ref{fig:divrot} shows that the two velocity gradient isotropy factors first slowly decrease according to the linear timescale $\Omega^{-1}$, reaching a moderate minimum of about 0.5 at the largest rotation rate. The time of maximum anisotropy for these quantities is close to that for $u_z' / u_x'$, although slightly lower, and here again the scaling with the linear timescale is broken during the increase at large time.  This plot shows that, in the final stage, the small scales are both three-dimensional and three-component, although not necessarily isotropic.

\begin{figure}
\begin{center}
\includegraphics[width=9.5cm]{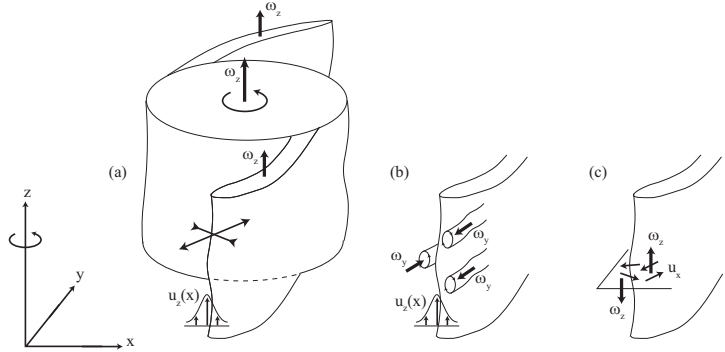}
\caption{Sketch showing thinning and instability of the vertical layers of vertical velocity advected by the horizontal flow. (a), Nearly vertical layer of ascending fluid, $u_z > 0$, strained in the vicinity of a large cyclone $\omega_z > 0$. (b), the layer becomes unstable, producing horizontal vortices. (c) These horizontal vortices produce random horizontal motion, and hence vertical vorticity of arbitrary sign.} 
\label{fig:sketch}  
\end{center}                                            
\end{figure}                                             

Assuming that the vertical velocity $u_z$ behaves as a scalar field passively advected by the large scale horizontal flow provides a qualitative explanation for the increase of $\omega_y' / \omega_z'$ at large time. As sketched in figure~\ref{fig:sketch}(a), a layer of ascending fluid $u_z>0$ in a horizontal strain field, for instance in the vicinity of a large vortex, is elongated along one direction and compressed along the other one, so it becomes thinner. In this process, $u_z$ is approximately conserved, but its horizontal gradient $\nabla_h u_z$ increases, producing horizontal vorticity $\omega_x$ and $\omega_y$ which may reach, and even exceed, the vertical vorticity $\omega_z$.

The increase of $\gamma_z'$, on the other hand, may be a consequence of the instability of those vertical plane jets. If the inertial timescale of the jets, $(\nabla_h u_z)^{-1} \simeq L_{33,1} / u_z'$, remains smaller than the dissipation timescale, $L_{33,1}^2 / \nu$, the jets may undergo shear instabilities, producing horizontal vortices (sketched in figure~\ref{fig:sketch}b), as suggested by the visualisations in figure~\ref{fig:p30_wy}(e-f). This condition is actually satisfied: The Reynolds number $Re_j$ based on those jets, defined as the ratio of the two timescales, writes
$$
Re_j = \frac{L_{33,1} u_z'}{\nu} = \frac{L_{33,1}}{L_f} \frac{u_z'}{u_x'} Re,
$$
where $Re$ is the instantaneous Reynolds number defined in Eq.~(\ref{eq:ReRo}). With $L_{33,1} / {L_f} \simeq 0.1$, $u_z' /u_x' \simeq 0.5$, and $Re$ ranging between 300 and 1300 in the final period of the decay (see figure~\ref{fig:ReRo}), one has $Re_j \simeq 10-10^2$, which is actually sufficient for a shear instability to develop. Little influence of the background rotation is expected on this shear instability, since the vertical velocity is unaffected by the Coriolis force (the resulting instability pattern, involving horizontal velocity, may however be affected by the rotation). The resulting wavy jets break the vertical invariance of $u_z$, thus producing vertical strain $\gamma_z'$ of the order of the vorticity $\omega_y'$, in agreement with figure~\ref{fig:divrot}.

All these results suggest that the flow structure at large time is fully three-dimensional and three-component, with isotropy factors (\ref{eq:isofac}) close to that of 3D isotropic turbulence, although the large scales are highly anisotropic, as described by the ordering of the integral scales (\ref{eq:order}).


\section{Cyclone-Anticyclone asymmetry}
\label{sec:vor}

\subsection{Dynamics of the cyclones and anticyclones}

\begin{figure}
\begin{center}
\includegraphics[width=9.5cm]{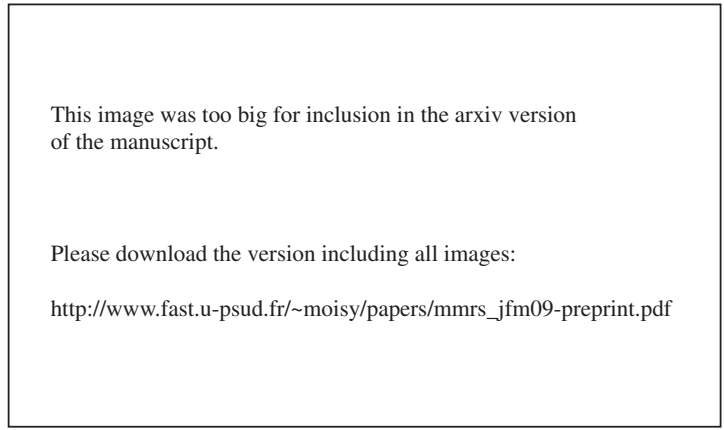}
\caption{Sequence of 6 snapshots of the velocity and vertical vorticity fields $\omega_z$ measured in a horizontal plane $(x,y)$ at mid-height for $\W = 0.20$ rad.s$^{-1}$. The imaged area is 1.3 m
$\times$ 1.3 m, representing 4.6\% of the tank section. The tank rotation is anticlockwise. Positive and negative vorticity indicate cyclones (in red) and anticyclones (in blue) respectively. The color range is normalized by the rms $\omega_z' = \langle \omega_z^2 \rangle_{w,y,e}^{1/2}$ computed for each time.}
\label{fig:wz}
\end{center}
\end{figure}

We finally turn to the structure of the vertical vorticity field in the rotating case, focusing on the issue of the cyclone-anticyclone asymmetry. The dynamics of the horizontal flow is illustrated by the 6 snapshots in figure~\ref{fig:wz} for $t>t^*$ (see also the Movie 2). At the beginning of the decay, the vorticity field consists in small scale disordered fluctuations (Fig.~\ref{fig:wz}a,b), which gradually evolve into a complex set of tangled vortex sheets and vortices (Fig.~\ref{fig:wz}c). A set of well defined, nearly circular, cyclones gradually emerges and separates from the turbulent background (Fig.~\ref{fig:wz}d).  Anticyclones are also encountered, but they are weaker and less compact than the cyclones.  Those anticyclones do not appear to be specifically unstable compared to the cyclones, so that the asymmetry seems to originate essentially from the enhanced vortex stretching of cyclonic vorticity.

At large time, the size of the cyclone grows, and merging of cyclones are frequently encountered, as illustrated by figure~\ref{fig:wz}(e). At the same time, a background of small scale symmetrical vorticity fluctuations appears (Fig.~\ref{fig:wz}e,f). At the end of the decay, the flow essentially consists in those small-scale symmetric fluctuations, advected by the large-scale, mostly cyclonic, motions.

\subsection{Growth of vorticity skewness}

\begin{figure}
\begin{center}
\includegraphics[width=6.5cm]{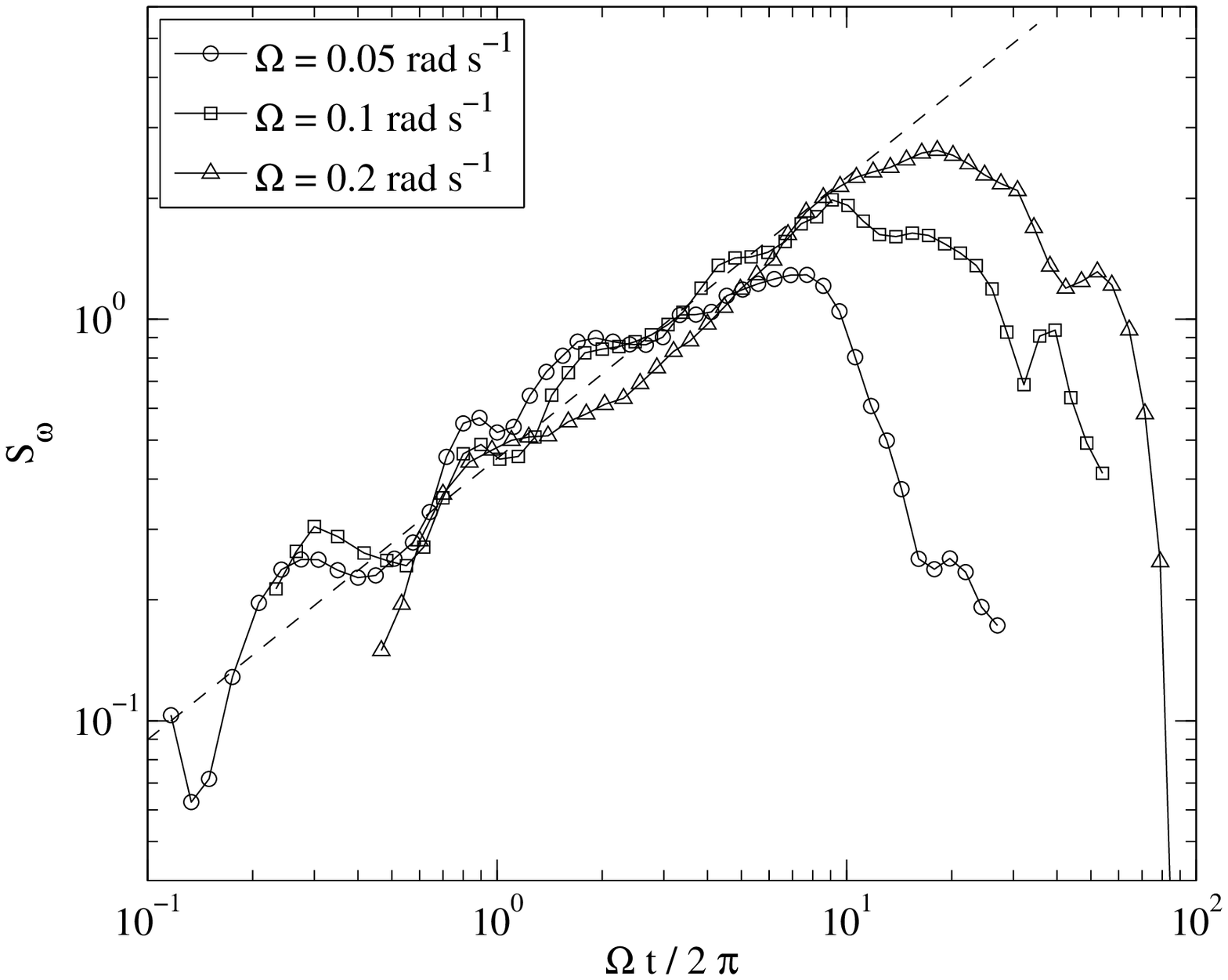}
\includegraphics[width=6.5cm]{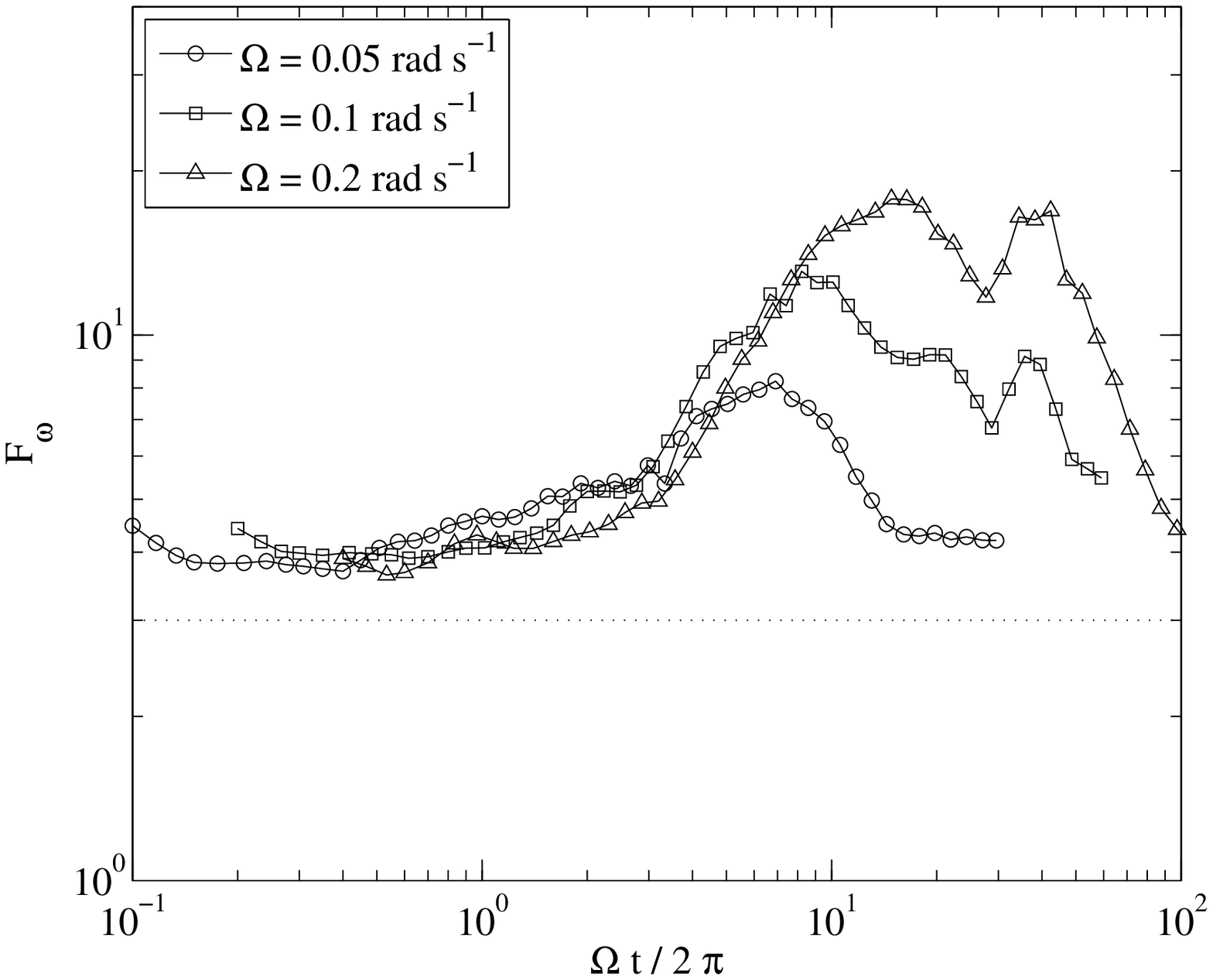}
\caption{Vorticity skewness  $S_\omega$, and (b) Vorticity flatness $F_\omega$, as a function of the number
of tank rotation $\Omega t/2\pi$. In (a), the dashed
line shows the fit $0.45 (\W t / 2\pi)^{0.7}$.
In (b), the  dashed line indicates the
value $F_\omega=3$ corresponding to a Gaussian field.} 
\label{fig:Sw}  
\end{center}                                            
\end{figure}

The gradual structuration of the vorticity field can be described by the vorticity skewness and flatness factors,
$$
S_\omega = \frac{\langle \omega_z^3 \rangle}{\langle \omega_z^2
\rangle^{3/2}}, \qquad F_\omega = \frac{\langle \omega_z^4 \rangle}{\langle \omega_z^2 \rangle^{2}},
$$
(where the brackets denote horizontal and ensemble average), which are plotted in Fig.~\ref{fig:Sw}. Both $S_\omega$ and $F_\omega$ show a growth followed by a decrease. In the growth regime at early time, both quantities collapse when plotted as a function of the number of tank rotations $\Omega t / 2 \pi$, indicating that the build-up of vortex structures is a linear mechanism. Note that the residual oscillations visible at small times are associated to the large scale IW flow (see Sec.~\ref{sec:meanflow}), of period $\Omega t / 2 \pi = 1/2$.  On the other hand, as for the isotropy factors, this rescaling with $\Omega^{-1}$ no longer holds during the decrease of $S_\omega$ and $F_\omega$ at large time.

For $t<t^*$, the vorticity skewness $S_\omega$ is essentially zero, within an uncertainty of $\pm 10^{-1}$. For $t>t^*$, it grows according to the power law
$$
S_\omega \simeq 0.45 \left( \frac{\Omega t}{2 \pi} \right)^{0.7},
$$
which is in remarkable agreement with the one reported by Morize \etal (2005), both concerning the exponent and the numerical pre-factor. Although in both experiments turbulence is generated by the translation of a grid, the details of the geometry differ in a number of respects: here the grid velocity is normal to the rotation axis, and the aspect ratio is significantly lower ($h / L_y = 0.25$ instead of 1.3). The collapse of $S_\omega$ for the two experiments is a clear indication of a generic behaviour of this quantity (Morize \etal 2006b).

The peak values of $S_\omega$, between 1.5 and 3 for increasing $\Omega$, are significantly larger than those obtained in the experiments of  
Morize \etal (2005) and Staplehurst \etal (2008), and in the DNS of Bokhoven \etal (2008). The corresponding peaks of $F_\omega$, between 8 and 18, are much larger than usually measured in non-rotating turbulence at similar Reynolds number (see, e.g., Sreenivasan \& Antonia, 1997), an indication of the strong concentration of vorticity in the core of the cyclones.


\subsection{The decay of vorticity skewness at large time}

The peaks of $S_\omega$ and $F_\omega$ occur at the time corresponding to the return to isotropy of the small scales, as identified in Sec.~\ref{sec:rti}. For larger times, when the flow consists mostly in isolated large-scale cyclones, $S_\omega$ decreases back to 0, while $F_\omega$ recovers values around 4, similar to the begining of the decay. The important scatter in the decay is due to the very limited sampling: At large times, the number of strong vortices per unit of imaged area becomes less than 1, so the statistics become very sensitive to events of vortices entering or leaving the field.

There is no general agreement concerning the decrease of $S_\omega$ at large time. It was attributed to confinement effects by Morize \etal (2005), namely the non-linear Ekman pumping of the cyclonic vortices (Zavala Sans\'on \& van Heijst 2000). Linear pumping, which is valid in the limit of $\omega \ll 2\Omega$, should imply identical decay of vorticity moments of both cyclonic and anticyclonic vorticity, and should have therefore no effect of $S_\omega$. This suggestion was motivated by the fact that the time $t_{max}$ of maximum $S_\omega$ was approximately following the Ekman time scale, $t_{max} \simeq 0.1 h (\nu \Omega)^{-1/2}$. Fitting the times of maximum skewness for the present data would actually give similar values, although the spread of the maximum of $S_\omega$ prevents from a real check of the $\Omega^{-1/2}$ scaling. However, the fact that the Ekman pumping is shown to have no significant effect in the present experiment (see Sec.~\ref{sec:aniso}) rules out this interpretation.  The role of the confinement in the decay of $S_\omega$ is also questioned by the numerical data of Bokhoven \etal (2008), who have reported a decrease of $S_\omega$ at large times in a homogeneous turbulence with periodic boundary conditions, and hence with no Ekman pumping. Note that no decrease of $S_\omega$ was reported in numerical simulation of Bourouiba \& Bartello (2007) and in the experiment of Staplehurst \etal (2008),  probably because of their limited temporal range (the measurements of the latter were restricted to 3 tank rotations, whereas the decrease of $S_\omega$ starts typically after 10 rotations here and in Morize \etal 2005).


The explanation we propose here for the decrease of $S_\omega$ relies on the fact that vorticity is a small-scale quantity, whereas the cyclone-anticyclone asymmetry is defined by structures of increasing size as time proceeds. In particular, it is observed that the flow outside the cyclones is not smooth, but is made of small scale vorticity fluctuations, approximately symmetric.
These vorticity fluctuations originate from the instabilities of the vertical shear layers strained by the horizontal large-scale flow described in Sec.~\ref{sec:rti}. The horizontal vortices resulting from this instability induce an horizontal straining flow (fig.~\ref{fig:sketch}b), which may itself be unstable and produce vertical vorticity of random sign at small scale (fig.~\ref{fig:sketch}c). Accordingly, there is a possibility for the vorticity skewness to return to zero at large time, although the large scale field remains dominated by a set of big cyclones.

\begin{figure}
\begin{center}
\includegraphics[width=8cm]{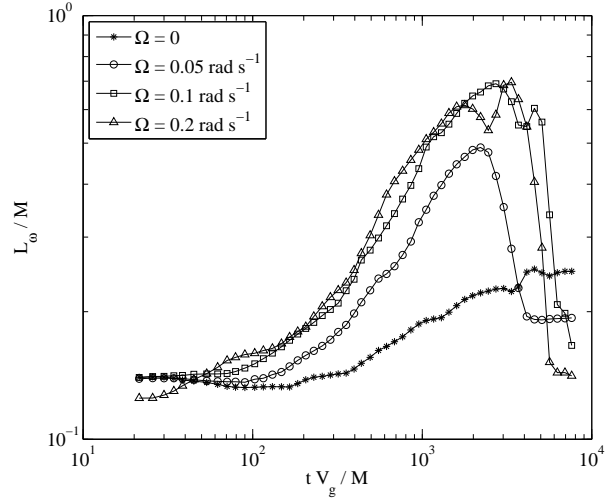}
\caption{Time evolution of the horizontal integral scale of vertical vorticity, $L_\omega$.}
\label{fig:lwvst}
\end{center}
\end{figure}

The characteristic size of the vortical structures may be estimated from the horizontal integral scale of the vertical vorticity, $L_\omega = (L^{\omega}_{33,1} + L^{\omega}_{33,2})/2$. Here the integral scales for the vorticity are defined similarly as those for the velocity, by modifying Eq.~(\ref{eq:LCex}) as
\begin{equation}
L^{\omega}_{33, \beta} = \int_0^{r^*}  \frac{\langle \omega_z ({\bf x}, t) \omega_z ({\bf x} + r {\bf e}_\beta, t) \rangle}{\langle \omega_z^2 \rangle} \, dr,
\label{eq:LCwex}
\end{equation}
where the truncation scale $r^*$ has the same meaning as in Eq.~(\ref{eq:LCex}). Figure~\ref{fig:lwvst} shows the characteristic increase and decrease of $L_\omega$ when rotation is present, whereas it monotonically increases in the absence of rotation. The decrease occurs at $t V_g / M \simeq 2000$ for all rotation rates, which coincides with the sharp decrease of $L_{33,1}$  (figure~\ref{fig:Laab}) and the isotropy factors (figure~\ref{fig:divrot}), and reaches values of order $0.2 M \simeq 30$~mm, similar to those found for $L_{33,1}$. This is a confirmation that the vertical vorticity field is dominated, at large time, by the small scale fluctuations induced by the instabilities of the vertical shear layers.

\begin{figure}
\begin{center}
\includegraphics[width=9.5cm]{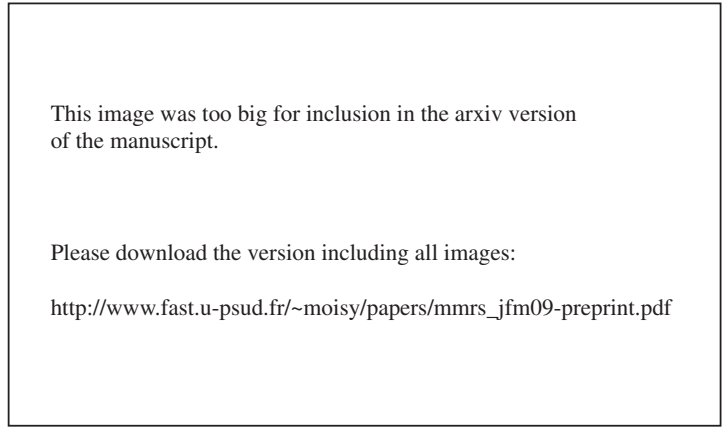}
\caption{Sequence of four $\omega_z$-snapshots in the horizontal plane $(x,y)$ at large time, showing the advection of the small-scale symmetric vorticity by the large scale horizontal motion ($\Omega = 0.10$~rad~s$^{-1}$, $t \simeq 64 M/V_g$). Each image is separated by $20$~s~$=T/3$, and the field of view is 1.3~m$\times$~1.3~m. The two ellipses track some arbitrary vorticity pattern in time. The angular velocity of the cyclonic structure in the bottom-left corner (white circle) is $\Omega_{c} \simeq 0.009$~rad~s$^{-1}$, corresponding to a local Rossby number of $\Omega_{c} /\Omega = 0.09$.} 
\label{fig:stripwz}  
\end{center}                                            
\end{figure}

The role of the symmetric small scale vorticity fluctuation in the decrease of $S_\omega$ must be addressed very carefully, because the vorticity field deduced from the PIV may obviously be affected by measurement noise. 
The curl of this PIV noise being essentially symmetric, it would yield a trivial reduction of $S_\omega$. This is a delicate issue, because the scale $L_\omega$ of those vorticity fluctuation is only slightly larger than the PIV resolution (Sec.~\ref{sec:pivres}). However, the temporal coherence of those small scale fluctuations advected by the large scales is evident at the end of the movie 2, whereas PIV noise would generate vorticity patterns essentially uncorrelated in time. Those vortical fluctuations cannot originate from the apparent particle displacements associated to the free surface disturbances (see \S.~\ref{sec:fsd}), which would yield a purely irrotational displacement field.  The temporal coherence of the flucutations may also be inferred from the 4 snapshots in figure~\ref{fig:stripwz}, where sets of arbitrary chosen vorticity patterns (marked in dashed ellipses) can be easily tracked in time, confirming that they are essentially advected by the large-scale motions.

\subsection{Skewness of the filtered vorticity field}

In order to check more precisely the influence of the measurement noise on $S_\omega$, we have computed $S_\omega$ from the filtered velocity ${\tilde {\bf u}}$ obtained by convolution of ${\bf u}$ with a Gaussian kernel of size $r_f$,
$$
{\tilde {\bf u}}(x,y, t ; r_f) = \int \! \! \! \int {\bf u}(x',y',t) \frac{1}{\sqrt{2\pi} r_f} e^{-\frac{(x-x')^2+(y-y')^2}{2 r_f^2}} dx' dy'.
$$
In practice the integral is restricted to a square area of size $6 r_f$.

\begin{figure}
\begin{center}
\includegraphics[width=6.5cm]{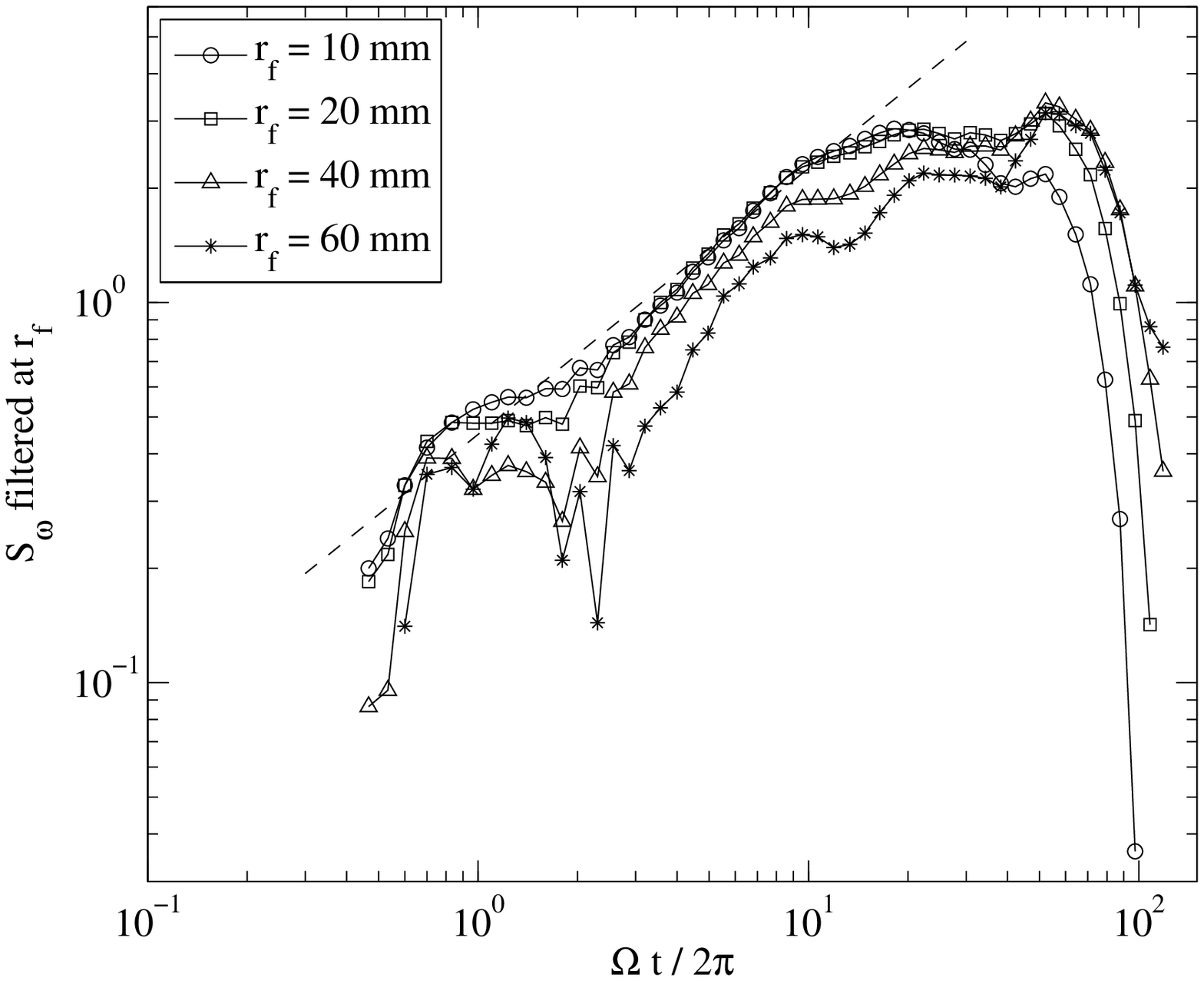}
\includegraphics[width=6.5cm]{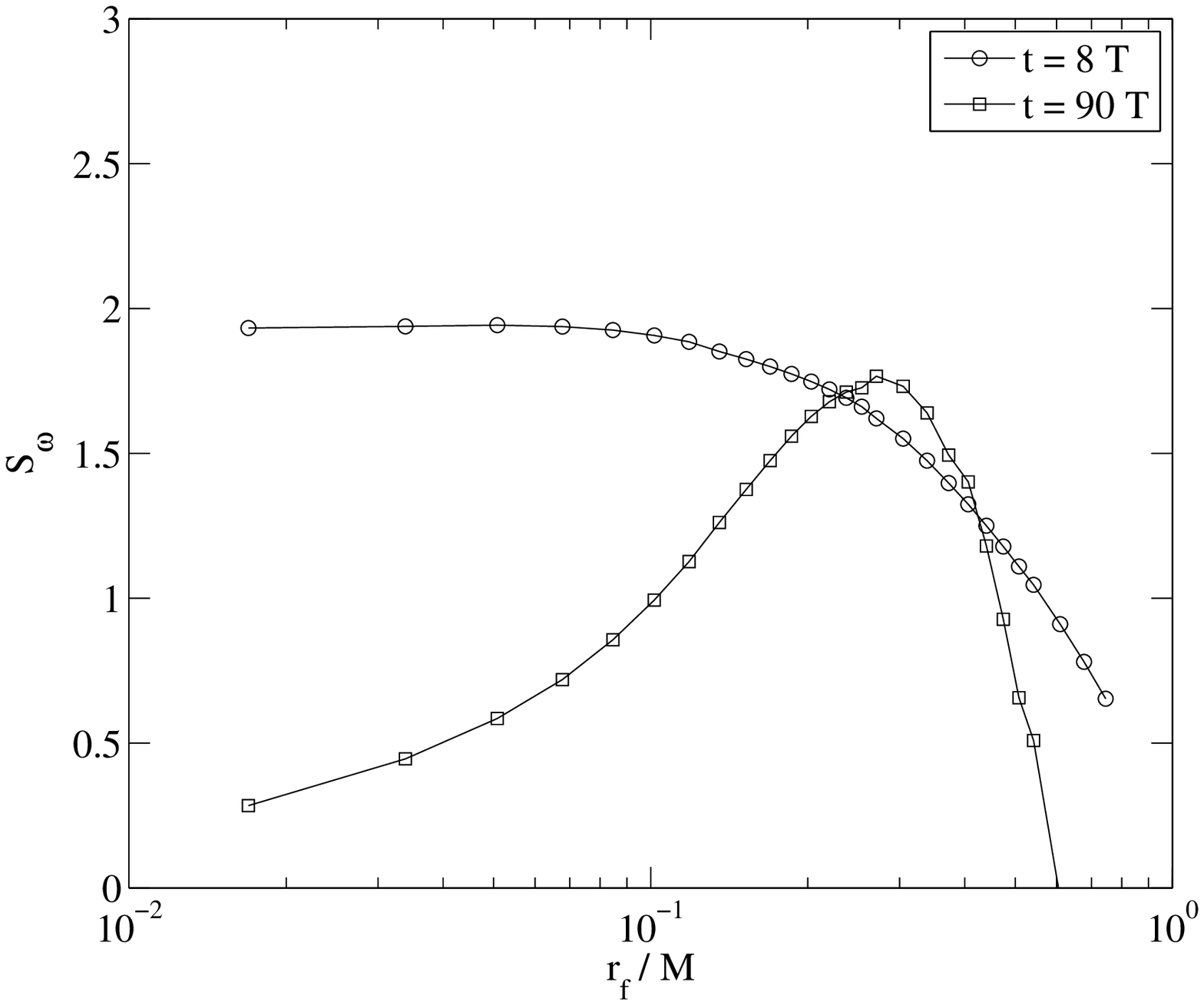}
\caption{(a), Time evolution of the skewness of the filtered vorticity field, $S_{\tilde \omega}$, for different filter size $r_f$, for $\Omega = 0.20$~rad~s$^{-1}$. (b), Vorticity skewness as a function of the filter size, at times $t=8T$ (before the peak of $S_\omega$) and $t=90T$ (after the peak).}
\label{fig:swvsf}
\end{center}
\end{figure}

The time evolution of the skewness of the filtered vorticity, $S_{\tilde \omega}$, is shown in figure~\ref{fig:swvsf}(a) for various filter sizes $r_f$, in the case $\Omega = 0.20$~rad~s$^{-1}$. In the growth regime, increasing the filter leads to a decrease of $S_\omega$, showing that the vorticity asymmetry is essentially contained at the smallest scale. In this situation, although the measured $S_\omega$ may underestimate the actual one because of the finite resolution of the PIV measurement, the vorticity skewness truly reflects the cyclone-anticyclone asymmetry at the smallest scales. On the other hand, after the peak of $S_\omega$, the ordering of the curves is reversed, so that filtering the vorticity field now increases the skewness, showing that now the asymmetry is carried by vortices at larger scales. This is consistent with figure~\ref{fig:stripwz}, where a large cyclone containing small-scale vorticity fluctuations is shown (white circle). However, although the peak of $S_\omega$ is shifted to larger times, a decrease is still observed. The effect of the filtering is further illustrated in figure~\ref{fig:swvsf}(b), where $S_{\tilde \omega}$ monotonically decreases as $r_f$ is increased at $t = 8 T$, whereas it shows a non-monotonic behavior at $t=90T$. Interestingly, in this latter case, the maximum of $S_{\tilde \omega}$ provides a rough estimate of the size of the vortices responsible for the cyclone-anticyclone asymmetry at this time.

One may conclude that, although $S_\omega$ provides a suitable description of the vortex asymmetry in the growth regime, when the characteristic size of the vortices corresponds to the diffusive scale (the 'Kolmogorov scale' modified by the rotation), it is no longer appropriate as the vortex size grows at larger time, in which case $S_\omega$ is strongly reduced by the small scale symmetric vorticity fluctuations. This does not imply, however, that filtering at even larger scales would totally inhibit the decrease of $S_\omega$, since other physical mechanisms, such as vortex pairing or diffusion, may also produce a reduction of $S_\omega$.

\section{Conclusion}
\label{sec:concl}

The present experiment aims to focus on the transition at Rossby number $Ro \simeq O(1)$ which occurs in the course of the decay of grid turbulence, initially approximately homogeneous and isotropic, in a rotating frame. Emphasis is given on the energy decay, anisotropy growth and asymmetry between cyclonic and anticyclonic vertical vorticity.

The different steps of the decay can be summarised as follows:

\begin{enumerate}

\item During the first 0.4 tank rotation (between 25 and 100~$M/V_g$),
the instantaneous Rossby number $Ro$ is larger than 0.25 and turbulence is not affected by the background rotation. Once the large scale mean flow and waves are properly subtracted, the turbulent energy follows the classical decay law $t^{-6/5}$ of isotropic unbounded turbulence.

\item After 0.4 tank rotation, $Ro < 0.25$ and the first effects of the rotation are triggered. Provided the grid Rossby number is large enough, the energy decay in  this regime is found to be compatible with the $\Omega^{3/5} t^{-3/5}$ law proposed by Squires \etal (1994), which is based on the assumption of energy transfers governed by the linear time scale $\Omega^{-1}$.
The horizontal flow becomes strongly correlated along the vertical direction, with a saturation of the integral scale $L_{11,3}$ by the vertical confinement.
Both the large-scale isotropy factor $u_z' / u_x'$ and the small-scale ones $\omega_y' / \omega_z'$ and $\gamma_z' / \omega_z'$ depart from their isotropic value, on the linear timescale $\Omega^{-1}$, although reaching only moderate value of about 0.5. A cyclone-anticyclone asymmetry develops by preferential vortex stretching of the cyclonic vorticity, and is well described by a power law growth of the vorticity skewness as $S_\omega \propto (\Omega t)^{0.7}$, consistent with the previous findings of Morize \etal (2005).

\item Finally, for $t > 2000 M/V_g$ (corresponding to 10-30 tank rotations), the vertical layers of vertical velocity, essentially advected by the large-scale horizontal flow, becomes thinner and prone to shear instabilities. These instabilities produce small scale spanwise vorticity, resulting in an apparent return to isotropy, which breaks the scaling with the linear timescale. However, the flow structure remains strongly anisotropic, as revealed by the characteristic ordering of the integral scales. A remarkable consequence of the instability of those vertical shear layers is that it re-injects horizontal velocity disturbances, and hence vertical vorticity with random sign, at small scales. Those symmetric vorticity fluctuations appear as a small-scale noise, which contributes significantly to the reduction of the vorticity skewness $S_\omega$, although the large scale vortices still remain preferentially cyclonic.

\end{enumerate}


The mechanism presented here provides an explanation for the decrease of $S_\omega$ at large time, which is observed here and in other configurations (Morize \etal 2005; Bokhoven \etal 2008). It must be noted that, although $S_\omega$ provides a suitable description of the vortex asymmetry in the regime (b), when the vortices responsible for the asymmetry are at small scale, it is no longer appropriate as the vortex size becomes larger, in the regime (c), when the flow consists in small-scale symmetric vorticity fluctuations superimposed to large scale quasi-horizontal motions. A more suitable statistical quantity, based for instance on the low-pass filtered vorticity or the transverse velocity increments, should provide a better description of the cyclone-anticyclone asymmetry in this regime.

Finally, the present results suggest that the initial conditions have a critical importance in the asymptotic state of decaying rotating turbulence, and in particular in the cyclone-anticyclone asymmetry. For initial isotropic turbulence, as is approximately produced in the wake of a grid, 
the initial vertical fluctuations, which represents 1/3 of the initial turbulent kinetic energy, is temporarily stored by the horizontal quasi-2D motions and plays no role in its dynamics, except at large time when it is released (and dissipated) directly at small scale via the shear instabilities of the vertical layers, resulting in a shortcut of the energy cascade. The effects of this energy re-injection at small scale is probably enhanced in the experiments of Morize \etal (2005), in which the grid was translated vertically instead of horizontally, resulting in a stronger decrease of the vorticity skewness at large time (Morize \etal 2006b). On the other hand, a rather different situation is expected for decaying rotating turbulence starting from strictly two-dimensional initial conditions, as in the experiments of Longhetto \etal (2002) and Praud \etal (2006), in which no vertical velocity is produced by the translation of a rake instead of a grid. In this latter situation, the vorticity skewness should reach significantly larger values, although a decay at large time may still be observed but for other reasons, such as vortex pairing or diffusion.

\begin{acknowledgments}

We gratefully acknowledge H. Didelle, S. Viboud, and A. Aubertin for experimental help, and C. Cambon, S. Galtier for fruitful discussions. This work was supported by the ANR grant no. 06-BLAN-0363-01 ``HiSpeedPIV''.

\end{acknowledgments}

\end{document}